\documentclass[12pt]{article}
\usepackage[top=2.75cm, bottom=2.75cm, left=2.75cm, right=2.75cm]{geometry}
\usepackage{color, amsthm}
\usepackage[centertags,reqno]{amsmath}
\usepackage{amssymb,amscd}
\usepackage[utf8]{inputenc}
\usepackage{hyperref,graphics,epsfig}
\usepackage{titlesec}
\usepackage{appendix}
\usepackage{microtype}

\hypersetup{
 pdftitle={Equivariant Rozansky-Witten invariants and TFT},    
    pdfauthor={Johan Kallen, Jian Qiu and Maxim Zabzine},
    pdfnewwindow=true,      
    colorlinks=true,       
    linkcolor=black,          
    citecolor= black,        
    filecolor= black,      
    urlcolor= black           
}

\definecolor{Blue}{rgb}{0.3,0.3,0.9}

\numberwithin{equation}{section}
\numberwithin{equation}{section}
\renewcommand{\theequation}{\thesection.\arabic{equation}}

\def\bi{{\bar i}}
\def\bj{{\bar j}}
\def\bk{{\bar k}}

\def\+{{+\!\!\!+}}

\newcommand{\smalint}{{\Large\textrm{$\smallint$}}}
\newcommand{\beq}{\begin{equation}}
\newcommand{\eeq}[1]{\label{#1}\end{equation}}
\newcommand{\bs}{\begin{split}}
\newcommand{\es}{\end{split}}
\newcommand{\ie}{{\it i.e.\ }}

\begin{document}

\setcounter{page}{0}
\newcommand{\inv}[1]{{#1}^{-1}} 
\renewcommand{\theequation}{\thesection.\arabic{equation}}
\newcommand{\bea}{\begin{eqnarray}}
\newcommand{\eea}{\end{eqnarray}}
\newcommand{\BS}{\boldsymbol}
\newcommand{\nn}{\nonumber}
\newcommand{\SF}{\mathsf}
\def\bra{{\langle}}
\def\ket{{\rangle}}

\thispagestyle{empty}
\begin{flushright} \small
UUITP-35/10\\
NSF-KITP-10-132
 \end{flushright}
\smallskip
\begin{center} \Large
{\bf Equivariant Rozansky-Witten classes and TFTs}
  \\[12mm] \normalsize
{\bf Johan~K\"all\'en$^a$, Jian Qiu$^b$ and Maxim Zabzine$^a$} \\[8mm]
 {\small\it
     ${}^a$Department of Physics and Astronomy,
     Uppsala university,\\
     Box 516,
     SE-751\;20 Uppsala,
     Sweden\\
     \vspace{.3cm}
     ${}^b$I.N.F.N. and Dipartimento di Fisica\\
     Via G. Sansone 1, 50019 Sesto Fiorentino - Firenze, Italy
}
\end{center}
\vspace{7mm}

\begin{abstract}
\noindent
We first construct the Rozansky-Witten model coupled to BF theory and Chern-Simons theory using the
Alexandrov-Kontsevich-Schwarz-Zaboronsky (AKSZ) method. Then we apply the machinery developed in some
earlier papers about AKSZ theories and characteristic classes to these concrete models: the BF-Rozansky-Witten
 model and the Chern-Simons-Rozansky-Witten
model. In the former case, we obtain characteristic
classes on the target hyperK\"ahler manifold equipped with a group action as a generalization of the original Rozansky-Witten
classes. We also give the prescription for similar classes associated with a holomorphic symplectic manifold and demonstrate the invariance of such classes explicitly.
\end{abstract}

\eject
\normalsize
\tableofcontents

\section{Introduction}

In \cite{Rozansky:1996bq} Rozansky and Witten described a three dimensional topological sigma model with the target
 space being  a hyperK\"ahler manifold $M$, and they showed that the partition function of the theory
  is a three-manifold invariant of finite type. The perturbative expansion of the Rozansky-Witten (RW) model
   gives rise to an interesting weight system which depends on the hyperK\"ahler manifold. The weights are labeled by
     trivalent graphs which are just Feynman diagrams of the field theory.
   Roughly speaking, if we choose a complex structure $J$ and a holomorphic symplectic form $\Omega$ on $M$,
    then we can use the Riemann curvature tensor of the hyperK\"ahler manifold to construct the vertex
  \bea
 V_{jkl}=dz^{\bi}R_{\bi j~l}^{~~n}\Omega_{nk}~\in~\Omega^{0,1}(M, T^* \otimes T^* \otimes T^*)~,\nn
 \eea
  where we have used the complex coordinates for $J$.  Taking a trivalent graph $\Gamma$ with $2n$ vertices
  we can contract  the vertices $V$ with $\Omega^{-1}$ according to the graph.  For example, the following
   graph\\
   \\
   \begin{figure}[h]
\begin{center}
\includegraphics[bb=0 0 113 51,width=.8in]{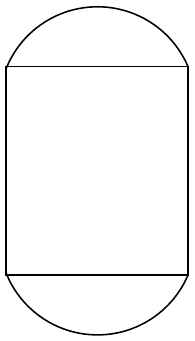}
\end{center}
\end{figure}\\
 corresponds to
\bea
V_{i_1 i_2 i_3} V_{j_1 j_2 j_3} V_{l_1 l_2 l_3} V_{k_1 k_2 k_3}  \Omega^{i_1 j_1} \Omega^{i_2 j_2} \Omega^{j_3 l_3}
 \Omega^{l_1 k_1} \Omega^{l_2 k_2} \Omega^{k_3 i_3}~,
\eea
which is a differential form in $\Omega^{0,4}(M)$.  For a general trivalent graph $\Gamma$ with $2n$ vertices
    the resulting form will be in $\Omega^{0,2n}(M)$, and  this form is $\bar{\partial}$-closed.
      Indeed it should be understood as
     an element of the Dolbeault cohomology group $H_{\bar{\partial}}^{0,2n}(M)$, and
      we refer to this class as a RW characteristic class. The crucial properties of a
      RW class associated to $\Gamma$ is that it depends on $\Gamma$ only through its homology class and it
       is invariant under deformations of the hyperK\"ahler metric.  Since its discovery in 1996
       many different aspects of the RW classes and invariants have been studied, and in fact
        the  RW classes can be defined for any holomorphic symplectic manifold \cite{Kontsevichformal, kapranov-1997}.
 We refer the reader to  \cite{sawon-phd} for a nice mathematical summary of RW theory.
 Recently in \cite{Qiu:2009zv, Qiu:2009rf} a treatment of the RW model within the Batalin-Vilkovisky (BV) formalism
  has been developed. The BV framework offers
  very elegant and natural explanations of many properties of  RW theory.  Actually, there is a canonical relation between
  a wide class of 3-dimensional TFTs and characteristic classes of graded (super) manifolds.
  The RW theory is just one particular manifestation of this generic phenomena.

In the present work our goal is  to construct the equivariant version of RW-classes for hyperK\"ahler
  manifolds  with a compatible group action.  We address this problem as physicists by constructing appropriate
   extensions of the original RW model and by studying  their perturbative expansion.
 The extension of the RW model we consider is a ``gauged" version of the RW model. Namely, we show that if the holomorphic
   symplectic manifold admits a Hamiltonian group action then we can couple the RW model to BF-theory or, upon some
   additional condition on moment map, to Chern-Simons (CS) theory.
   Our analysis is very similar to the ideas presented in \cite{Qiu:2009rf} and the present work can be regarded as
   a concrete illustration for the general ideas advertised in \cite{Qiu:2009rf}.
  Previously the RW model coupled to CS theory was written down and studied by Kapustin and Saulina in \cite{Kapustin:2009cd}
   and it is related to the topological twist of  the Gaiotto-Witten model \cite{Gaiotto:2008sd}.
By assuming that there is a holomorphic moment
map on the hyperK\"ahler manifold, one can enlarge the BRST transformations of the RW model with extra terms involving the
group action. Here we will construct the same model with
a more streamlined and uniform approach known as the AKSZ construction. Due to the need to make the BRST transformations
nilpotent, an extra requirement was imposed on the moment map. This requirement restricts the applicability of the model.
On the other hand, the RW model coupled to BF theory is more liberal, and also as we shall see, the perturbative expansion
of such models only has a finite number of terms, in sharp contrast to the CS-RW case.

 Let us briefly summarize the perturbative results coming from BF-RW theory.
  The RW model can be formulated on a manifold $M$ with a holomorphic symplectic form $\Omega$, and we assume that
    there is a group acting holomorphically and preserving $\Omega$. Assuming that the group action is Hamiltonian with holomorphic moment map $\mu_A$
   we can write down the RW model coupled to BF theory whose 3-valent vertex $V_{ijk}$ is
  \bea
 V_{jkl}=dz^{\bi}R_{\bi j~l}^{~~n}\Omega_{nk}+A^{A} \nabla_{j}\nabla_{k}\partial_{l}\mu_{A}
+ A^Ak_A^i R_{ij~l}^{~\,n} \Omega_{nk} ~\in ~ \Omega^{0,1}(M, T^* \otimes T^* \otimes T^*)
\oplus \wedge^1 {\mathbf g}^*~,\label{vertex-BF-CS}
 \eea
  where we assume symmetrization in $j,k, l$ and $k_A^i = \partial_j \mu_A\Omega^{ji}$.
 Here ${\mathbf g}^*$ is the dual of the Lie algebra of the Lie group acting on $M$ and $A^A$ is basis element in ${\mathbf g}$
  of odd degree.  In order to write down $V$ we have to pick up a connection on $M$ compatible with $J$ and $\Omega$
   (which always exists).  For a hyperK\"ahler manifold we can just pick the Levi-Civita connection for the hyperK\"ahler metric
    and the last term in the vertex $V_{jkl}$ is identically zero.
  By taking a trivalent graph $\Gamma$ with $2n$ vertices we contract the $V$'s using $\Omega^{-1}$ in the same
   fashion as in the RW story. The resulting object
   \bea
 c_\Gamma \in \bigoplus\limits_{p+q=2n} \Omega^{0,p}(M)\otimes \wedge^q {\mathbf g}^*\label{observable}
\eea
  is closed under the following differential
\bea Q=\bar\partial+A^Ak_A^i\partial_i-\frac12f_{AB}^{~~~C}A^AA^B\frac{\partial}{\partial A^{C}}~.\nn\eea
  We can define the corresponding cohomology group $H^{2n}_Q(M)$
  and understand $c_\Gamma$ as a cohomology class.
  The class satisfies the following properties:\\

$\bullet$ it is invariant under a deformation of the connection (hyperK\"ahler metric) on $M$\\

$\bullet$ it depends on the graph $\Gamma$ only through its graph homology class.\\

  These classes give rise to a weight system for the perturbative expansion of the BF-RW model.
At each order $\hbar^n$ of perturbation, there is a collection of 3-valent
graphs as dictated by Wick's theorem. The partition function (without integrating over zero modes) has
 the form
 \bea
 \sum\limits_{\Gamma} b_\Gamma c_\Gamma~,\label{partfunc}
 \eea
  where the $b_\Gamma$'s depend on the three manifold and are the same as in the RW model.  This expansion has a finite number of
   terms and the number of terms depends on the dimensionality of $M$.
  Many nice properties of the expansion (\ref{partfunc})
   follow from manipulations using the BV machinery, which we will review later.
In sections \ref{AKSZ-general},\ref{perturbative},\ref{char-classes}, we restrict ourselves to $M$ being hyperK\"ahler to simplify the formulae. The generalization to any holomorphic symplectic manifold with a Hamiltonian group action is given in section \ref{math}.
\\
\\
 A similar analysis can be performed for the CS-RW model. However, there are additional requirements on the moment maps
 and the perturbative expansion continues to all orders.  There still exists an interpretation in terms of appropriate characteristic classes.
\\
\\
The paper is organized as follows: In  section \ref{AKSZ-general}
 we first construct the RW, BF-RW and CS-RW models using a systematic approach within the BV framework.
  In particular, the CS-RW model is identical with the one constructed by Kapustin and Saulina in \cite{Kapustin:2009cd}.
 In section \ref{perturbative}, we show that the new construction allows us to do computations with superfields and that leads to a tremendous simplification in the perturbative computation and we obtain the same results as those obtained with the more traditional component approach.
  We present the Feynman rules for the RW, BF-RW and CS-RW models.
 In section \ref{char-classes} we give an interpretation of the perturbative partition function for these models in terms of
  appropriate characteristic classes.    We show heuristicaly why these classes are closed and
 independent of certain data.
  In section \ref{math} some formal and mathematical issues are collected and
  discussed. In particular we discuss the general case of a holomorphic symplectic manifold with a Hamiltonian group action.
 Section \ref{end} gives a summary of the results and some outlook.

\section{Systematic Construction of the RW, BF-RW and CS-RW Model}
\label{AKSZ-general}

\subsection{Review of the AKSZ construction of the RW Model}
\label{Review-RW}

Topological field theories have since long been a very effective tool in producing and studying
topological invariants.  Typically for a TFT whose fields $\BS\Phi$ are mappings
\bea
\BS\Phi:~~\BS{\Sigma}~\to~\BS{{\cal M}}~,\nn
\eea
the action of the model can usually be written down using totally canonical data on
$\BS{\Sigma}$ and $\BS{{\cal M}}$. In particular, no metric data is used. As a result, the expectation values of gauge invariant operators
are expected to depend only on these canonical data and nothing else.
 Here we name $\BS{\Sigma}$ and $\BS{{\cal M}}$ as the source and target space respectively, both may be super (graded) manifolds.
 In \cite{Alexandrov:1995kv} a  conceptually clear approach for constructing topological field theories was introduced.
 Besides  its easy-to-use  feature, this approach meshes well with the BV formalism, and therefore offers one the ability to discuss the gauge dependence of the theory in a way that the BRST formalism cannot. This approach is known under the name
 of the Alexandrov-Kontsevich-Schwarz-Zaboronsky (AKSZ) construction. Using geometrical
  data on $\BS{\Sigma}$ and $\BS{{\cal M}}$, the AKSZ approach gives
    a canonical solution of the classical master equation within the BV-framework.
For a general discussion and examples of the AKSZ approach, we refer to
 \cite{Cattaneo:2001ys,Cattaneo:2009zx,Qiu:2009zv,Ikeda:2010vz,2007LMaPh79-143R} and more references therein.
  We will not review the general construction here nor will we use the
language of graded manifolds in any essential way, but rather focus on some specific 3D TFTs.

Our source manifold is $\BS{\Sigma} = T[1]\Sigma$, where $\Sigma$ is a 3-manifold
with coordinate $x^a$. We can assemble forms on $\Sigma$ into superfields by introducing a formal odd (degree 1) variable $\theta^a$ which transforms as $dx^a$. Therefore a polyform on $\Sigma$ is just a function of $\theta^a,x^a$ \footnote{The convention here is slightly different from \cite{Qiu:2009rf,Qiu:2009zv} in that now the product of superfields is just the wedge product.}:
\bea f(x,\theta)=f(x)+\theta^af_a(x)+\frac12\theta^a\theta^b f_{ab}(x)+\frac16\theta^a\theta^b\theta^c f_{abc}(x)~.\nn\eea
Here $f(x)$ is a 0-form, $f_a(x)$ is a 1-form etc. Often we denote $f(x)$ as $f_{(0)}$, $dx^a f_a(x)$ as $f_{(1)}$ and so on (also for the zero form component if no confusion is likely, we even drop the subscript ${}_{(0)}$). The multiplication of superfields directly corresponds to the wedge product and the de Rham differential is $D=\theta^a\partial_a$.

Our target manifold ${\cal M}$ is a degree 2 graded symplectic manifold, which means that there is a symplectic structure $\omega$
 of degree $2$. Written in local Darboux coordinates
 $$\omega =\frac{1}{2} \omega_{AB}~d\Phi^A\wedge d\Phi^B~,$$
 where  $\omega_{AB}=-(-1)^{|\Phi^A||\Phi^B|}\omega_{BA}$ and $\Phi^A$ is the coordinate on ${\cal M}$ with
 $|\Phi^A|$ denoting its degree.
   The BV space will be the space of superfields $\BS{\Phi}^A=\Phi^A(x,\theta)$, \ie space of maps\footnote{This space requires
   proper mathematical definition, e.g. see \cite{Cattaneo:2009zx} for a discussion. However this is not relevant to our considerations
    and we work constantly in local coordinates.}, Maps$(T[1]\Sigma,{\cal M})$.
There is an odd symplectic form on the BV space:
\bea
\omega_{BV}=\frac{1}{2}\int d^6z~ \omega_{AB}~\delta \BS{\Phi}^{A}\delta \BS{\Phi}^{B}~,\label{BV_symp}
\eea
where $d^6z=d^3xd^3\theta$ and if any function is written in bold then we assume that
  it is a superfield. With such an odd symplectic form, one can define a naive Laplacian
\bea
\Delta=\int d^3x~ (\omega^{-1})^{AB}(-1)^{|\Phi^A|}\Big(\frac{\partial}{\partial \Phi^A_{(3)}}\frac{\partial}{\partial \Phi^B_{(0)}}
+\frac{\partial}{\partial \Phi^A_{(1)}}\frac{\partial}{\partial \Phi^B_{(2)}}\Big)~.\label{Laplacian}
\eea
The Laplacian induces an odd Poisson bracket in the BV space in the following way
\bea
 \{f,g\}=(-1)^{|f|}\Delta(fg)-(-1)^{|f|} (\Delta f)g-f(\Delta g)~,~~f,g\in \textrm{Maps}(T[1]\Sigma,{\cal M})~.\label{BV_bracket}
 \eea
There is an important identity
\bea
\Big\{\int d^6z~ f(\BS{\Phi}),\int d^6z~ g(\BS{\Phi})\Big\}=-\int d^6z~ \{f,g\}(\BS{\Phi})~,\label{property_Laplacian}
\eea
which relates the bracket in the BV space to the bracket in the target space ${\cal M}$.

The path integral is defined not over all BV space, but rather over a Lagrangian
submanifold in the BV space on which (\ref{BV_symp}) vanishes.
 The choice of this Lagrangian submanifold is called the \emph{gauge fixing}. The following key statement
is  about gauge invariance in the BV framework: if a function ${\cal O}$ is annihilated by the Laplacian (\ref{Laplacian}),
then the integral of ${\cal O}$ is invariant under small changes of the Lagrangian submanifold ${\cal L}$. Furthermore the integral
of something $\Delta$-exact is zero. This says that the action of any theory must satisfy $\Delta e^S=0$, or equivalently
$\Delta S+1/2\{S,S\}=0$. This equation is known as the
quantum master equation. In fact, usually the classical master equation $\{S,S\}=0$ and the quantum one are fulfilled simultaneously.

The main innovation of \cite{Alexandrov:1995kv} is to encode the classical master equation (which is on the space of mappings) into a single function $\Theta$ on the target space ${\cal M}$ with the property $\{\Theta,\Theta\}=0$ (this is non-trivial only when $\Theta$ is of odd degree). The property of $\Theta$ would ensure
\bea
 \{\int d^6z~ \BS{\Theta},\int d^6z~ \BS{\Theta}\}=-\int d^6z~\{\BS{\Theta},\BS{\Theta}\}=0~.\nn
 \eea
The term $\Theta$ will be used as the interaction term for the theory, we can also write a kinetic term. Suppose
there is a Liouville form $\Xi_A d\Phi^A$ such that  $d\Xi=\omega$, then the kinetic term is
\bea S_{kin}=\int d^6z~ \BS{\Xi}_A (\BS{\Phi})~D\BS{\Phi}^A.\nn\eea
Notwithstanding the fact that $\Xi$ may only exist locally, the kinetic term is well defined if $\partial\Sigma=\emptyset$. We will
constantly use the Darboux coordinates where $\Xi_A=\Phi^B\Omega_{BA}$.

The full action is then given by
\bea
S=S_{kin}+S_{int}=\int d^6z ~ \Big(\frac12\BS{\Phi}^A\Omega_{AB}D\BS{\Phi}^B+\BS{\Theta}\Big)~.\nn
\eea
The action automatically satisfies $\{S,S\}=0$ and $\Delta S=0$.

The hard  and non-canonical part is naturally the gauge fixing. The rough idea is that, sometimes there are quite 'natural' choices of ${\cal L}$,
but the resulting action has too much residue gauge symmetry rendering the path integral ill defined. So we perturb ${\cal L}$ a bit in order to fix those residue gauge symmetry.

Let us take a look at the Rozansky-Witten model and see how to arrive at it from our approach.
Let $M$ be a complex manifold admitting a holomorphic symplectic form $\Omega$. Denote the target space of the RW model as the
 graded manifold $\mathcal{M}_{RW} \equiv T^{*0,1}[2] (T^{*0,1}[1]M)$, which is locally parameterized by the following: $X^{i},X^{\bar i}$ coordinates of $M$;
$p_{\bar i}$ degree 2 fiber coordinates in the anti-holomorphic cotangent
direction; $v^{\bar i}$ degree 1 fiber coordinate of $T^{0,1}M$; $q^{\bar i}$ degree 1 fiber coordinate of $T^{*0,1}M$.
 We choose the following symplectic form  of degree $2$ on $\mathcal{M}_{RW}$
\bea
\omega_{RW}=dp_{\bi}d X^{\bi}+dq_{\bi}dv^{\bi}+\frac{1}{2}\Omega_{ij}dX^{i}dX^{j}~,
\label{omegarw}\eea
where $\Omega_{ij}$ is the holomorphic 2-form on $M$ and we allocate degree $2$ for $\Omega$ (alternatively we can
 introduce formal parameter of degree $2$ in front of $\Omega$).
 The natural choice of Hamiltonian function of degree $3$ is
\bea
\Theta=-p_{\bi}v^{\bi}~,\nn
\eea
which satisfies $\{\Theta,\Theta\}=0$, and it acts as $\bar\partial$ on functions of $f(X,v)$. With these data, the BV action is given by
\bea
S_{RW}=\int{d^{6}z~\left(\BS{p}_{\bi}D\BS{X}^{\bi}+\BS{q}_{\bi}D \BS{v}^{\bi}+\frac{1}{2}\Omega_{ij}\BS{X}^{i}D\BS{X}^{j}-\BS{p}_{\bi}\BS{v}^{\bi}\right)}~.
\label{rwaction}
\eea
 By construction this action satisfies the classical master equation on the BV space
 $\textrm{Maps}(T[1]\Sigma,{\cal M}_{RW})$.

 The only data we used so far is the holomorphic symplectic structure on $M$. Now we will discuss the gauge fixing.
 For the sake of clarity and simplicity of formulae we specialize now
 to the hyperK\"ahler case, then $\Omega_{ij}$ is covariantly constant with respect to the Levi-Civita connection $\Gamma^{j}_{ik}$ and $\Gamma^{\bj}_{\bi\bk}$.
  However, we would like to stress that this restriction is not essential and  everything can be carried out for a generic holomorphic
   symplectic manifold (see section \ref{math}).
Skipping some algebra, the Lagrangian submanifold is given by the set of conditions
\bea &&{\tilde p}_{\bar i(0)}={\tilde p}_{\bar i(1)}={\tilde p}_{\bar i(2)}
=X^i_{(2)}=X^i_{(3)}=\BS{q}_{\bar i}=0~,\nn\\
&&
{\tilde p}_{\bar i(3)}=\frac16\Omega_{lj}R_{i\bi~k}^{~~j}X_{(1)}^l\wedge X_{(1)}^i\wedge X_{(1)}^k~.\label{RW_gauge_fix}\eea
Note that we have defined a new variable
\beq
\tilde{\BS{p}}_{\bi}=\BS{p}_{\bi}+\Gamma^{\bj}_{\bi\bk}\BS{q}_{\bj}\BS{v}^{\bk}~,\nn
\eeq{skskaa}
which transforms as a tensor in contrast to $\BS{p}_{\bi}$. Also, the components of fields are now defined using covariant
derivatives, e.g. $v^{\bar i}_{ab}=1/2[\nabla_{\theta^b},\nabla_{\theta^a}]\BS{v}^{\bar i}|_{\theta=0}$. The reader may
 consult \cite{Qiu:2009zv} for more details.

 One may check that this set of conditions
set $\omega_{BV}$ to zero. Evaluated on this Lagrangian submanifold, one finds the action
\beq
S_{RW}=\frac12\int{d^{3}x\left(\Omega_{ij}X^{i}_{(1)}\wedge d^{\nabla}X^{j}_{(1)}-\frac{1}{3}R^{~~~i}_{k\bk ~j}X^{k}_{(1)}\wedge \Omega_{li}X^{l}_{(1)}\wedge X^{j}_{(1)}v^{\bk} \right)}~,
\eeq{}
where $d^{\nabla}X_{(1)}^{i}=dX_{(1)}^{i}+\Gamma^{i}_{jk}dX^{j}_{(0)}X^{k}_{(1)}$.
We notice that there is no kinetic term for the fields $X_{(0)}^{i},X_{(0)}^{\bi},v^{\bar i}$,
and the kinetic term for $X_{(1)}$ is not invertible, since the de Rham operator has an infinite dimensional kernel.
We use the freedom of deforming the Lagrangian submanifold to obtain a nice quadratic term.
The safest way is to deform every field by $\delta\phi=\{\Psi,\phi\}$,
where $\phi$ is any field in the theory and $\Psi$ is a cleverly chosen function. We notice that this kind of deformation will maintain the condition $\omega_{BV}|_{{\cal L}+\delta{\cal L}}=0$,
since the deformation is generated by a Hamiltonian vector field. For the RW model we choose
\beq
\Psi=-\frac{1}{2}\left(g_{i\bj}X^{i}_{(1)}\wedge *dX_{(0)}^{\bj}\right)~.\nn
\eeq{sjs2992011}
Applying $\Psi$ to deform the Lagrangian submanifold defined by (\ref{RW_gauge_fix}), we get the complete action
\bea
&&S_{RWgauged}=\frac{1}{2}\int d^{3}x\Big(g_{i\bj}dX^{i}_{(0)}\wedge *dX^{\bj}_{(0)}-g_{i\bj}X^{i}_{(1)}\wedge *d^{\nabla}v_{(0)}^{\bj}\nn\\
&&\hspace{2.7cm}+\Omega_{ij}X^{i}_{(1)}\wedge d^{\nabla}X^{j}_{(1)}-\frac{1}{3}R^{~~~i}_{k\bk ~j}X^{k}_{(1)}\wedge \Omega_{li}X^{l}_{(1)}\wedge X^{j}_{(1)}v^{\bk} \Big)~,
\label{rwactionfinal}\eea
which agrees with the theory constructed originally by Rozansky and Witten in \cite{Rozansky:1996bq}.

According to the general prescription of BV-AKSZ \cite{Alexandrov:1995kv}, the BRST transformations for the theories are always obtained by calculating $\delta_{BRST}\phi=\{S,\phi\}|_{{\cal L}}$. For the RW model we find
\bea
&&\delta X^{\bar{i}}=v^{\bar{i}}~,~~\delta X^i=0~,\nn\\
&&\delta X_{(1)}^{ {i}}=dX^{ {i}}~,\nn\\
&&\delta v^{\bar{i}}=0~.\nn\eea
Note that $\{S,S\}=0$ off shell, but the BRST transformation is the restriction of $\{S,\cdot\}$ onto ${\cal L}$ and may only close on-shell
 in general. In the RW model the transformations close off-shell.

\subsection{BF-RW  Model}

In this section, we will couple the RW model to gauge fields. Let us assume that the
 holomorphic symplectic manifold $(M, J, \Omega)$ underlying the RW model admits an action of a
  Lie group ${\mathbf G}$.
  Let us assume that this action preserves the complex structure $J$ and the holomorphic symplectic structure $\Omega$.
   At infinitesimal level the action is realized by the vector fields $k_A$ such that the Lie brackets are given by
\beq
[k_{A},k_{B}]=f_{AB}^{~~~C}k_{C}~,
\eeq{liebracket}
 and we assume that ${\cal L}_{k_A} J=0$ and ${\cal L}_{k_A} \Omega=0$.  Thus in complex coordinates
  we can write  $k^{\mu}_{A}\partial_\mu =k^{i}_{A}(z) \partial_i +k^{\bi}_{A}(\bar{z}) \partial_{\bi}$.
  Finally let us assume that the action is Hamiltonian with respect to $\Omega$, \ie
  there exist a holomorphic moment map $\mu_{A}$ defined by
\beq
\begin{split}
\partial_{j}\mu_{A}&=k^{i}_{A}\Omega_{ij}~,
\end{split}
\eeq{momentmapdef}
 which is equivariant
 \bea
\{\mu_{A},\mu_{B}\}_{\Omega}=f_{AB}^{~~~C}\mu_{C}~,
\label{equivariant}
\eea
where $\{~,~\}_{\Omega}$ is Poisson bracket with respect to the holomorphic symplectic form $\Omega$.
 Next we introduce the graded manifold  ${\cal M}_{BF-RW} \equiv ({\mathbf g}^*[1]\otimes {\mathbf g}[1])\times {\cal M}_{RW}$ with the  coordinates of ${\mathbf g}^*[1]\otimes {\mathbf g}[1]$ being $B_A$ and $A^A$ of degree 1 and ${\cal M}_{RW}$
  was defined in the previous subsection. This space is equipped  with an even  symplectic form of degree $2$
\bea \omega_{BF-RW}=\omega_{RW}+\delta B_A\delta A^A~,\nn
\eea
 where $\omega_{RW}$ is defined in (\ref{omegarw}).
 The Hamiltonian of degree $3$ is defined as follows
\beq
\Theta=-p_{\bi}v^{\bi}+\frac{1}{2}f_{AB}^{~~~C}A^{A}A^{B}B_{C}{\color{black}{+\mu_{A}A^{A}}}~,\nn
\eeq{}
where $\mu_A$ is of degree 2 (since $\Omega$ was assumed to be degree 2)
The corresponding master equation is satisfied
\beq
\{\Theta,\Theta\}=f_{AB}^{~~~D}f_{DE}^{~~~F}A^{A}A^{B}A^{E}B_{F}+\left(\{\mu_{A},\mu_{B}\}_{\Omega}-f_{AB}^{~~~C}\mu_{C}\right)A^{A}A^{B}=0~,\nn
\eeq{}
 since the first term vanishes due to the Jacobi identity and second due to
 the equivariance of the moment maps. Also we use the fact that the moment map $\mu_A$ is holomorphic, \ie
  $\bar{\partial} \mu_A =0$.

 With this data it is straightforward to apply the AKSZ construction. On the space
 $\textrm{Maps}(T[1]\Sigma,{\cal M}_{BF-RW})$ there is a BV-bracket defined by the following odd
  symplectic structure
 \beq
\omega_{BV}=\int{d^{6}z~\left(\delta \BS{B}_{A} \delta \BS{A}^{A}+\delta \BS{p}_{\bi}\delta \BS{X}^{\bi}+\delta \BS{q}_{\bi}\delta \BS{v}^{\bi}+\frac{1}{2}\Omega_{ij}\delta \BS{X}^{i}\delta \BS{X}^{j}\right)}~.
\eeq{omegacsrwkdkdk}
 The master action defining the BF-RW model is thus
\bea
&&S_{BF-RW}=\int d^{6}z~\Big(\BS{B}_{A}D\BS{A}^{A}+ \BS{p}_{\bi}D\BS{X}^{\bi}+\BS{q}_{\bi}D \BS{v}^{\bi}\nn\\
&&\hspace{3.3cm}+\frac{1}{2}\Omega_{ij}\BS{X}^{i}D\BS{X}^{j}+\frac{1}{2}f_{AB}^{~~~C}\BS{A}^{A}\BS{A}^{B}\BS{B}_{C}-\BS{p}_{\bi}\BS{v}^{\bi}{\color{black}{+\mu_{A}\BS{A}^{A}}}\Big)~.
\label{BFRWaction}\eea
 This model is defined for any holomorphic symplectic manifold with a Hamiltonian action of a group.

Now to proceed to the gauge fixing, we again specialize to the hyperK\"ahler case and we choose the Lagrangian submanifold for the RW sector in the same way as in (\ref{RW_gauge_fix}).  For the gauge sector, we use the metric
 on $\Sigma_3$ to Hodge decompose the differential forms on $\Sigma_3$. That is to say, we decompose any superfield $\BS{A}$ (and similarly for $\BS{B}$) according to
\bea \BS{A}=\BS{A}^h+\BS{A}^e+\BS{A}^c=\BS{A}^h+\theta^a\partial_a\BS{r}-\nabla^a \partial_{\theta^a}\BS{s},\nn\eea
where $c,e,h$ stands for coexact, exact and harmonic; $\BS{r,s}$ are some superfields and $-\nabla^a \partial_{\theta^a}$ is just $d^{\dagger}$ written in the super language. If one decomposes the above equation into components, one gets exactly the usual Hodge decomposition.

The symplectic BV-form is thus decomposed as
\bea \int d^6z~\delta \BS{B}_A\delta\BS{A}^A=\int d^6z~\delta (\BS{B}_A)^c\delta(\BS{A}^A)^e+
\delta (\BS{B}_A)^e\delta(\BS{A}^A)^c+\delta (\BS{B}_A)^h\delta(\BS{A}^A)^h~.\nn\eea
We do the gauge fixing by choosing the following Lagrangian submanifold:
 $(\BS{B}_A)^e=(\BS{A}^A)^e=0$ and
 for the harmonic sector we set $A^h_{(3)}=A^h_{(2)}=B^h_{(3)}=B^h_{(2)}=0$. We find the complete action for the gauge-fixed BF-RW model to be
 {\small\beq
\bs
S_{BFRW}&=S_{RWgauged}+S_{BF}+S_{new} \\
S_{RWgauged}&=\frac{1}{2}\int\left(\Omega_{ij}X^{i}_{(1)}\wedge D^{A}X^{j}_{(1)}-\frac{1}{3}R^{~~~i}_{k\bk~j}X^{k}_{(1)}\wedge \Omega_{li}X^{l}_{(1)}\wedge X^{j}_{(1)}v^{\bk}\right) \\
S_{BF}&=\int\Big(\big(B_{(1)A}\wedge d A_{(1)}^{A}+\frac{1}{2}f_{AB}^{~~~C}A_{(1)}^{A}\wedge A_{(1)}^{B}\wedge B_{(1)C}\big)- B_{(2)C}\wedge\big(d A_{(0)}^{C}+f_{AB}^{~~~C}A^{A}_{(1)}A_{(0)}^{B}\big)\Big) \\
&\hspace{2cm} +A_{(2)}^{A}\wedge\big(-dB_{(0)A}-f_{AB}^{~~~C}A_{(1)}^{B}B_{(0)C}+f_{AB}^{~~~C}A_{(0)}^{B}B_{(1)C}\big)\Big) \\
S_{new}&=\int{\left( \partial_{i}\mu_{A}X^{i}_{(1)}\wedge A^{A}_{(2)}-\frac{1}{6}(\nabla_{i}\nabla_{j}\partial_{k} \mu_{A})X^{i}_{(1)}\wedge X^{j}_{(1)}\wedge X^{k}_{(1)}A^{A}_{(0)}\right)}~,
\end{split}
\eeq{blablabla}}
where $D^{A}X^{i}_{(1)}=d^{\nabla}X^{i}_{(1)}+(\nabla_{j}k^{i}_{A}) A_{(1)}^{A}\wedge X_{(1)}^{j}$. For the fields $A_{(0,1,2)},B_{(0,1,2)}$ in this expression, the gauge condition has been imposed even though we do not make explicit the superscript ${}^h$ or ${}^c$. This also explains the absence of $A_{(3)},B_{(3)}$.
  One can rewrite this action in more familiar form related to the Faddeev-Popov trick
   by introducing Lagrange multipliers which enforce the Lorentz
   gauge for one-forms
   and rewriting $2$-forms explicitly as $d^\dagger$ of another field.

One can also deform the gauge choice
by using $\Psi=-1/2~g_{i\bar j}X^i_{(1)}\wedge*D^AX^{\bar j}$ as we did in the previous subsection.
 The derivation of $S_{new}$ in (\ref{blablabla})  is very similar to the CS-RW model, and we will present
 some technical  details in the next subsection.

\subsection{CS-RW model}
\label{CSRWgaugefix}

In this subsection we construct the CS-RW model within the AKSZ approach and we will show that upon
 specific gauge fixing it is the same model discussed previously in  \cite{Kapustin:2009cd}
 within the BRST framework.

For the CS-RW model, let us define the target manifold as the following graded manifold
\beq
{\cal M}_{CS-RW} \equiv \mathbf{g}[1]\times \mathcal{M}_{RW}~.
\eeq{}
If the algebra ${\mathbf g}$ is equipped with an $ad$-invariant  metric $\eta_{AB}$ then on  ${\cal M}_{CS-RW}$
 there is a symplectic form of degree $2$
 \beq
\omega_{CS-RW} = \omega_{RW} + \frac{1}{2} \eta_{AB}\delta A^A \delta A^B~,
\eeq{ddkdkkd229}
 where $A^B$ are coordinates of degree $1$ on ${\mathbf g}$.   Let us assume as in the previous subsection
  that $M$ admits the holomorphic Hamiltonian action with moment maps $\mu_A$.
 Let us consider the following Hamiltonian function of degree $3$ on ${\cal M}_{CS-RW}$
\beq
\Theta=- p_{\bi} v^{\bi}+\frac{1}{6}f_{ABC} A^{A} A^{B} A^{C}+\mu_{A} A^{A}~,
\eeq{}
where $f_{ABC}=f_{AB}^{~~~D}\eta_{DC}$ and $\mu_A$ is of degree 2 (since $\Omega$ was assumed to be degree 2).
  In order to fulfill the master equation, we need
\beq
\{\Theta,\Theta\}=-\frac{1}{4}f_{AB}^{~~~C}f_{CDE} A^{A} A^{B} A^{D} A^{E}+\left(\{\mu_{A},\mu_{B}\}_{\Omega}-f_{AB}^{~~~C}\mu_{C}\right)
 A^{A} A^{B}{\color{black}{-\mu_{A}\mu_{B}\eta^{AB}}}=0~.
\eeq{}
Compared to the discussion in previous subsection, this gives us an extra constraint on the moment maps, namely
 $\mu_{A}\mu_{B}\eta^{AB}=0$.   Despite the fact that this additional condition looks exotic there are examples,
  see the discussion in \cite{Kapustin:2009cd}.

 Now on the space  $\textrm{Maps}(T[1]\Sigma,{\cal M}_{CS-RW})$ there is the following BV symplectic form
\beq
\omega_{BV}=\int{d^{6}z~\left(\frac{1}{2}\eta_{AB}\delta \BS{A}^{A} \delta \BS{A}^{B}+\delta \BS{p}_{\bi}\delta \BS{X}^{\bi}+\delta \BS{q}_{\bi}\delta \BS{v}^{\bi}+\frac{1}{2}\Omega_{ij}\delta \BS{X}^{i}\delta \BS{X}^{j}\right)}~.
\eeq{omegacsrw}
The action defining the CS-RW model is thus given by
\bea
&&S_{CSRW}=\int d^{6}z~\Big(\frac{1}{2}\eta_{AB}\BS{A}^{A}D\BS{A}^{A}+ \BS{p}_{\bi}D\BS{X}^{\bi}+\BS{q}_{\bi}D \BS{v}^{\bi}+\frac{1}{2}\Omega_{ij}\BS{X}^{i}D\BS{X}^{j}\nn\\
&&\hspace{3.3cm}+\frac{1}{6}f_{ABC}\BS{A}^{A}\BS{A}^{B}\BS{A}^{C}-\BS{p}_{\bi}\BS{v}^{\bi}+\mu_{A}\BS{A}^{A}\Big)~.
\label{CSRWaction}\eea
 This action automatically satisfies the master equation provided that there is a holomorphic Hamiltonian group
  action on the holomorphic symplectic manifold with the additional property\footnote{It is interesting to point out that the same condition
   appears in Gaiotto-Witten work \cite{Gaiotto:2008sd} and it is related to the Chern-Simons theory for super-Lie algebra.} on the moment maps  $\mu_{A}\mu_{B}\eta^{AB}=0$.

 Next let us discuss the gauge fixing of the present model.
Again, we specialize to the hyperK\"ahler case for simplicity.
We will now demonstrate that the above AKSZ model is indeed equivalent to the model obtained in \cite{Kapustin:2009cd}. The symplectic form of the RW sector and CS sector decouple, so we may choose the gauge fixing for the RW sector just as before, while for the CS sector, we set the exact part of each component to zero $A^e_{(1)}=A^e_{(2)}=A^e_{(3)}=0$ and also $A_{(3)}^h=A_{(2)}^h=0$. We obtain the standard CS action plus the
RW sector $\eqref{rwactionfinal}$ and a new part from expanding
\beq
S_{new}=\int{d^{6}z\left(\mu_{A} (\BS{X}^i) \BS{A}^{A}\right)}~.\nn
\eeq{330kdke2}
Performing the grassmanian integral, we find
{\small\bea
\int d^{3}x~\left(-\frac16(\nabla_{i}\nabla_{j}\partial_{k} \mu_{A})X^{i}_{(1)}X^{j}_{(1)}X^{k}_{(1)}A^{A}_{(0)}-\frac12\Omega_{ij}\nabla_{k}k^{i}_{A} X^{k}_{(1)}X^{j}_{(1)}A^{A}_{(1)}+ \partial_{i}\mu_{A}X^{i}_{(1)}A^{A}_{(2)}\right).\label{muA}\eea}
In general, if we have an action on the manifold which preserves the metric\footnote{It is important to stress that we need the condition
 ${\cal L}_{k_A} g=0$ only to match with the results from \cite{Kapustin:2009cd}.
 In further discussion of the construction of characteristic classes we do not have
  to assume any special properties of metric or connection under the group action.},
 the vector field generating the action satisfies the Killing equation $\nabla_{(\mu}k^A_{\nu)}=0$, where $k^A_{\nu}=\eta^{AB} g_{\nu\mu}k^{\mu}_B$. From this equation and the Bianchi identity for the Riemann tensor it follows that
\beq
\nabla_{\mu}\nabla_{\nu}k^A_{\rho}=-R_{\rho\mu~\nu}^{~~\lambda}k^A_{\lambda}~.
\eeq{}
On a hyper-K\"ahler manifold, with an action preserving the holomorphic symplectic form and the K\"ahler form, we can obtain a stronger relation for $k^{\bi}_{A}$. In \cite{Kapustin:2009cd} it was shown that under these circumstances we have
\beq
(\nabla_{k}\nabla_{j}\partial_{i} \mu_{A})X^{k}_{a}X^{j}_{b}X^{i}_{c}A^{A}_{(0)}=-\Omega_{kl}R^{~~~l}_{i\bk~j}X^{k}_{a}X^{j}_{b}X^{i}_{c}k^{\bk}_{A}A_{(0)}^{A}~.
\eeq{}
Using this relation together with $\eqref{muA}$ and $\eqref{rwactionfinal}$, we find the total action to be
{\small\bea
&&S=S_{CS}+S_{RW} +S_{new}~,\nn\\
&&S_{CS}=\int\Big(\frac{1}{2}\eta_{AB}A_{(1)}^{A}\wedge dA_{(1)}^{B}-\eta_{AB}A_{(2)}^{A}\wedge dA^{B}_{(0)}\nn\\
&&\hspace{3.5cm}{\color{black}{+\frac{1}{6}f_{ABC}A_{(1)}^{A}\wedge A_{(1)}^{B}\wedge A_{(1)}^{C}}}-f_{ABC}A^{A}_{(2)}\wedge A^{B}_{(1)}A^{C}_{(0)}\Big)~,\nn \\
&&S_{RW}=\frac{1}{2}\int\Big(\Omega_{ij}X^{i}_{(1)}\wedge d^{\nabla}X^{j}_{(1)}-\frac{1}{3}R^{~~~i}_{k\bk~j}X^{k}_{(1)}\wedge \Omega_{li}X^{l}_{(1)}\wedge X^{j}_{(1)}v^{\bk}\Big)~,\nn \\
&&S_{new}=\int\Big(\frac{1}{6}R^{~~~i}_{k\bk~j}X^{k}_{(1)}\wedge \Omega_{li}X^{l}_{(1)}\wedge X^{j}_{(1)}k_{A}^{\bk}A^{A}_{(0)}\nn\\
&&\hspace{3.5cm} -\frac{1}{2}\Omega_{ki}\nabla_{j}k^{k}_{A} X^{i}_{(1)}\wedge X^{j}_{(1)}\wedge A^{A}_{(1)}+ \partial_{i}\mu_{A}X^{i}_{(1)}\wedge A^{A}_{(2)}\Big)~,\eea}
where $d^{\nabla}X_{(1)}^{i}=d X^{i}_{(1)}+\Gamma^{i}_{jk}X^{j}_{(1)}d X^{k}_{(0)}$. We see that
 the combination $v_{(0)}^{\bi}-k^{\bi}_{A}A^{A}_{(0)}$ naturally appears and it is useful to introduce the field
 $\eta^{\bi}=v_{(0)}^{\bi}-k^{\bi}_{A}A^{A}_{(0)}$.
 Furthermore, we can define a "big" covariant derivative by $D^{A}X^{i}_{(1)}=d^{\nabla}X^{i}_{(1)}+\nabla_{j}k^{i}_{A} A_{(1)}^{A}\wedge X_{(1)}^{j}$. This derivative is covariant under both gauge transformations and changes of coordinates. With these definitions, the final form of the action is
\beq
\begin{split}
&S_{CS}+S_{RW}+S_{new}= \\
&S_{CS}+\frac{1}{2}\int{\left((\Omega_{ij}X^{i}_{(1)}\wedge D^{A}X^{j}_{(1)}-\frac{1}{3}R^{~~~i}_{k\bk~j}X^{k}_{(1)}\wedge \Omega_{li}X^{l}_{(1)}\wedge X^{j}_{(1)}\eta^{\bk}+ \partial_{i}\mu_{A}X^{i}_{(1)}\wedge A^{A}_{(2)}\right)}~.
\end{split}
\eeq{finalaction}
 We can analyze the BRST-transformations related to this
  gauge fixing in the same way as for the RW model.
 At the level of components, the BRST transformations are given by
\bea
&&\delta X^{\bi}_{(0)}=v_{(0)}^{\bi}~,~~~
\delta X^{i}_{(0)}=k^{i}_{A}A^{A}_{(0)}~,\nn \\
&&\delta X^{i}_{(1)}=dX^{i}_{(0)}-k^{i}_{A}A^{A}_{(1)}-(\partial_{j}k^{i}_{A})X^{j}_{(1)}A^{A}_{(0)}~,\nn\\
&&\delta A^{A}_{(0)}=-\frac{1}{2}f_{BC}^{~~~A}A^{B}_{(0)}A^{C}_{(0)} -\eta^{AB}\mu_{B}~,\nn\\
&&\delta A^{A}_{(1)}=dA_{(0)}^{A}+f_{BC}^{~~A} A^{B}_{(1)}A^{C}_{(0)}+\eta^{AB}\partial_{i}\mu_{B}X^{i}_{(1)}~,\nn \\
&&\delta v^{\bi}_{(0)}=0~.
\eea
In order to come in contact with \cite{Kapustin:2009cd}, let us rewrite using $\eta^{\bi}=v_{(0)}^{\bi}-k^{\bi}_{A}A^{A}_{(0)}$. We find
\beq
\delta\eta^{\bi}={\color{black}{-\partial_{\bj}k^{\bi}_{A}v^{\bj}_{(0)}A^A_{(0)}+\frac12f_{BC}^{~~A}A^B_{(0)}A^C_{(0)}k_A^{\bar i}
+\eta^{AB}k^{\bi}_{A}\mu_{B}=-\partial_{\bj}k^{\bi}_{A}\eta^{\bj}A^{A}_{(0)}+\eta^{AB}k^{\bi}_{A}\mu_{B}}}
\nn\eeq{sss}
with the help of $\eqref{liebracket}$. With the field $\eta^{\bi}$, the relevant changes to the BRST transformation are
\bea
&&\delta X^{\bi}_{(0)}=\eta^{\bi}+k^{\bi}_{A}A^{A}_{(0)}~,\nn \\
&&\delta\eta^{\bi}_{(0)}={\color{black}{-\partial_{\bj}k^{\bi}_{A}\eta^{\bj}A^{A}_{(0)}+\eta^{AB}k^{\bi}_{A}\mu_{B}}}~,\nn
\eea
while the rest stays the same. Comparing with section 2.3 in \cite{Kapustin:2009cd}, we find the exact same BRST transformations, and the same action. In order to get kinetic terms for $X_{(0)}$ we add the standard BRST-exact term
\beq
\delta\left(g_{i\bj}X^{i}_{(1)}\wedge * d^{A} X^{\bj}_{(0)}\right)~.
\eeq{}
This will also produce a kinetic term for the co-exact part of $X^{i}_{(1)}$ and for the scalar $\eta^{\bi}$, just like in the RW-model.

The authors of \cite{Kapustin:2009cd} also added the term
\beq
\delta\left(\sqrt{h}g_{i\bj} k^{i}_{A}\bar{\mu}_{B}\eta^{AB} \eta^{\bj}\right)~,
\eeq{}
where $h$ is the determinant of the metric on the source, and $\xi^{i}=k^{i}_{A}\bar{\mu}_{B}\eta^{AB}$.

We have now reproduced the action and BRST transformations obtained in \cite{Kapustin:2009cd} within the AKSZ framework. In \cite{Kapustin:2009cd} they have a general function $f(A)$ and a Lagrange multiplier field $B$ whose equation of motion enforces $f(A)=0$. Our gauge fixing corresponds to $f(A)=\nabla^{a}A^{A}_{a}$.

Next our goal is to study the perturbative partition function for RW, BF-RW and CS-RW models. The gauge fixing of the BV models we have discussed so far may be useful for the interpretation of the models, however it is not very practical for doing perturbative calculations.
In the next section we will discuss a gauge fixing for  the actions $\eqref{BFRWaction}$ and $\eqref{CSRWaction}$ at the level of superfields
 and this will allows us to have simple Feynman rules.
By general arguments presented in the coming sections, the results of the calculation will correspond to
 certain characteristic classes of some differential $Q$ on the target space.
  But before specializing to concrete models, we will go through some general features of the perturbation theory for 3D AKSZ models.

\section{Perturbation Expansion of the Models}
\label{perturbative}

The perturbative expansion of the models we have discussed gives invariants both for the source manifold $\Sigma$ and for the target manifold ${\cal M}$. To gain insight into these invariants, it pays off to reorganize the calculations slightly.
 We use the RW model as an example, after that we will realise that a more straightforward gauge fixing can make the computation much simpler.

The action for the RW model (\ref{rwactionfinal}) is
\bea &&S_{RW}=\frac{1}{2}\int d^{3}x\Big(g_{i\bj}dX^{i}_{(0)}\wedge *dX^{\bj}_{(0)}-g_{i\bj}X^{i}_{(1)}\wedge *d^{\nabla}v_{(0)}^{\bj}\nn\\
&&\hspace{2.7cm}+\Omega_{ij}X^{i}_{(1)}\wedge d^{\nabla}X^{j}_{(1)}
-\frac{1}{3}R^{~~~i}_{k\bk ~j}X^{k}_{(1)}\wedge \Omega_{li}X^{l}_{(1)}\wedge X^{j}_{(1)}v^{\bk} \Big)~.\nn\eea

Since the boson quadratic term is BRST-exact (recall that it comes from $-\delta\Psi$), the path integral localises on
the zero of the quadratic term $dX^i=dX^{\bar i}=0$, namely constant maps. One can therefore expand the coordinate as
$X^{i(\bar i)}(x)=\textsl{x}_0^{i(\bar i)}+\xi^{i(\bar i)}(x)$ and treat $\textsl{x}_0^{i,\bi}$ as parameters. Note, by nature of such an expansion the zero mode of $\xi$ should be set to zero by hand.
Using some counting arguments one realises that $v^{\bar i}$ never participates in the perturbation theory actively other than that it sets the exact component of $X^i_{(1)}$ to zero. Furthermore there are only 3-valent vertices (one can see \cite{Rozansky:1996bq} for the argument, but we will arrive at this conclusion later in a more transparent way).

The curvature term is a ready made 3-valent vertex, and there is another one arising from expanding the connection in the second term
\bea g_{i\bar i}\Gamma^{\bar i}_{\bar j\bar k}dX^{\bar j}v^{\bar k} \wedge * X^i_{(1)}\to
 \nabla_l(g_{i\bar i}\Gamma^{\bar i}_{\bar j\bar k})\xi^ld\xi^{\bar j}v^{\bar k}\wedge* X^i_{(1)}
= g_{i\bar i}R^{~~\bar i}_{l\bar j~\bar k}\xi^ld\xi^{\bar j}v^{\bar k}\wedge*X^i_{(1)}~.\label{V_3_2}\eea
There are also other 3-valent vertices, but they contain an excess of $\xi^i$ and will not contribute.

From the kinetic terms we have the propagators, which we denote as
\bea (\Omega^{-1})^{ij}H_{ab}(x_1,x_2)=\bra X_a^i(x_1)X_b^j(x_2)\ket~,~~
g^{i\bar j}G(x_1,x_2)=\bra \xi^i(x_1)\xi^{\bar j}(x_2)\ket~.\nn\eea
Things still look rather complicated, but a sleight of hand can make them easier. Define
\bea X_{ab}^i=-(\Omega^{-1})^{ij}g_{j\bar j}\sqrt{h}\partial_c\xi^{\bar j}\epsilon^c_{~ab}~,~~~ \textrm{or}~~~
X^i_{(2)}=-(\Omega^{-1})^{ij}g_{j\bar j}*d\xi^{\bar j}~.\nn\eea
Using this notation, the vertex (\ref{V_3_2}) becomes
\bea
R_{l\bar k~i}^{~~j}\xi^l\Omega_{jm}X^m_{(2)}v^{\bar k}X^i_{(1)}~,\nn
\eea
which is begging to be combined with the other 3-valent vertex into a superfield form.
Also notice
\bea \bra X^i_{(2)}(x_1)\xi^j(x_2)\ket=-(\Omega^{-1})^{ik}g_{k\bar k}*d_1\bra \xi^{\bar k}(x_1)\xi^j(x_2)\ket
=-(\Omega^{-1})^{ij}*d_1G(x_1,x_2)~,\nn\eea
which would be exactly the propagator between $X^i_{(2)}$ and $\xi^j$ had there been a kinetic term
$\Omega_{ij} X^i_{(2)}d\xi^j$.


Taking advantage of this observation, we can define a superfield
\bea \BS{\xi}^i=\xi^i+\theta^a(X_a^i)^c+\frac{1}{2}\theta^a\theta^bX^i_{ab}~,\nn\eea
and assemble the kinetic terms into the super form
\bea S_{kin}=\int d^6z~\frac12\BS{\xi}^i\Omega_{ij}D\BS{\xi}^j~,\nn\eea
the two vertices into a single term
\bea V=\int d^6z~\frac16R_{\bar ij~l}^{~~m}\Omega_{mk}\BS{\xi}^j\BS{\xi}^k\BS{\xi}^lv^{\bar i}~,\nn\eea
and finally the component propagators into a super propagator
\bea (\Omega^{-1})^{ij}G(x_1,\theta_1;x_2,\theta_2)=\bra\BS{\xi}^i(x_1,\theta_1),\BS{\xi}^j(x_2,\theta_2)\ket~.\label{propagator-xi}\eea
In this way, the perturbative expansion can be done with the simplified action
\bea &&\int D\BS{\xi}~\exp\int d^6z\Big(\frac12\BS{\xi}^i\Omega_{ij}D\BS{\xi}^j+\frac16R_{\bar ii~k}^{~~j}\BS{\xi}^i\Omega_{jl}\BS{\xi}^l\BS{\xi}^k  v^{\bar i}\Big)~.\label{simplified_action}\eea
In this formula only $\BS{\xi}$ participates in the perturbation theory, and from the way $\BS{\xi}^i$ is made,
it has only a co-exact part and therefore the kinetic term is invertible. The fields $v^{\bar i}$ are treated as parameters.

This simplification is perhaps already foreseen by the reader,
since the RW model can be constructed through the AKSZ construction which uses superfields. And from the BV action (\ref{rwaction}) we can get the above simplified action using another gauge choice, as follows. By looking at the action (\ref{rwaction}), one realizes that $\BS{p}_{\bar i}, \BS{X}^{\bar i}$ are merely spectator fields, since $p_{\bar i}$ appears linearly in the action. The same happens to $\BS{q}_{\bar i}$ and $\BS{v}^{\bar i}$. The only active field is $\BS{X}^i$, and we would like to choose a gauge fixing condition for this field so that a superfield computation is possible.

The idea is to use Hodge decomposition and decompose every field into harmonic, exact and co-exact components.
Obviously we can perform the Hodge decomposition at the level of superfields.
${\cal L}$ is then chosen to set to zero all exact components of the superfields. However, such a decomposition for the coordinate $\BS{X}^i$ is improper, since the coordinates do not have a linear structure. To fix it, we choose a basis point $\textsl{x}_0$ as before and a tangent vector $\xi^i$ at $\textsl{x}_0$. But the expansion cannot be the simple minded
$\BS{X}^i=\textsl{x}_0^i+\BS{\xi}^i$, but rather the geodesic exponential map (here we use a connection with purely holomorphic indices)
\bea
X^i\stackrel{\exp}{=}\textsl{x}^i_0+\xi^i-\frac{1}{2}\Gamma^i_{jk}\xi^j\xi^k+\cdots~
=\exp\big\{\xi^i\partial_{\textsl{x}^i_0}-\xi^i\xi^j\Gamma^k_{ij}\partial_{\xi^k}\big\}\textsl{x}_0^i~.\nn
\label{exp_map}\eea
The symplectic form (\ref{omegarw}) is pulled back by the exponential map to
\bea\exp^*\omega_{RW}=\delta{p}_{\bar
i}\delta{X}^{\bar i}+\delta{q}_{\bar
i}\delta{v}^{\bar
i}+\frac{1}{2}\delta{\xi}^i\Omega_{ij}\delta{\xi}^j-\delta{X}^{\bar
i}\delta{\Theta}_{\bar i}~,\label{uebusheform}\eea
where
\bea
\Theta_{\bar i}(\xi)=\sum_{n=3}^{\infty}\Theta_{\bar in}~,
~~~\Theta_{\bar in}=\frac{1}{n!}\nabla_{l_4}\cdots\nabla_{l_n}R_{\bar il_1~l_3}^{~\,~k}\Omega_{kl_2}\,\xi^{l_1}...\xi^{l_n}~.\label{def_Theta}\eea
If the curvature of the manifold is of type $(1,1)$ then the covariant derivatives above commute with each other.
The action (\ref{rwaction})  is pulled back as
\bea &&\exp^*S=\int d^6z\ (\BS{p}_{\bar i}+\BS{\Theta}_{\bar
i})D\BS{X}^{\bar
i}+\frac{1}{2}\BS{\xi}^i\Omega_{ij}D\BS{\xi}^j+\BS{q}_{\bar
i}D\BS{v}^{\bar i}-\BS{p}_{\bar
i}\BS{v}^{\bar i}~.\nn\eea%
Since now the momentum dual to $X^{\bar i}$ is $p_{\bar
i}+\Theta_{\bar i}$, we should make the change of variable
$p_{\bar i}\to p_{\bar i}+\Theta_{\bar i}$:
\bea
&&\exp^*S=\int d^6z\ \BS{p}_{\bar i}D\BS{X}^{\bar
i}+\frac{1}{2}\BS{\xi}^i\Omega_{ij}D\BS{\xi}^j+\BS{q}_{\bar
i}D\BS{v}^{\bar i}-\BS{p}_{\bar i}\BS{v}^{\bar
i}+\BS{\Theta_{\bar i}}\BS{v}^{\bar i}~.\nn
\eea%
Note that $-p_{\bar i}v^{\bar i}+\Theta_{\bar i}v^{\bar i}$
generates the vector field $(v^{\bar i}\bar\partial_{\bar
i}+v^{\bar i}\{\Theta_{\bar i},\cdot\})$.

We see that the first term in $\Theta$ is just the interaction term in (\ref{simplified_action}) and we will argue later that only this first term matters. $\Theta$ satisfies the Maurer-Cartan equation
\bea \bar\partial_{[\bar i}\Theta_{\bar j]}=-\{\Theta_{\bar i},\Theta_{\bar j}\}\label{key_id}~.\eea

Let us now forget about the spectator fields and focus on the action\footnote{Note this action fails the master equation by a
$\bar \partial$-exact term, due to (\ref{key_id}).}
\bea \int d^6z~\frac{1}{2}\BS{\xi}^i\Omega_{ij}D\BS{\xi}^j+\BS{\Theta_{\bar i}}v^{\bar i}~.\nn\eea
Now this theory lives in a linear space and Hodge decomposition can be carried out. We set the exact part of $\BS{\xi}$ to zero. This clearly defines a Lagrangian submanifold, and the action thus restricted is
\bea
\int d^6z~\frac{1}{2}(\BS{\xi}^i)^c\Omega_{ij}D(\BS{\xi}^j)^c+\Theta_{\bar i}(\BS{\xi}^c)v^{\bar i}~,\label{RW_xi_c}
\eea
where $v^{\bar i}$ is merely a parameter.

Several other observations can further simplify the formalism. These observations apply to
any 3 dimensional AKSZ model. First by using integral by part, we get
\beq
\int{d^{6}z~\BS{\xi}^c(z)\BS{\xi}^{c}(z)}=0~,\nn\eeq{}
where we have thrown away a surface term. So the smallest non-trivial vertex will be 3-valent. Moreover, for a term with $n$ vertices, the number of propagators $p$ will be given by
\beq
p=\frac{3}{2}n~.
\eeq{p3halvan}
This is simply due to the fact that the propagator is quadratic in the $\theta$'s, and we must saturate all the $\theta$ integration. Since $p$ is integer, the number of vertices $n$ must be even.

Let now $k_i$ be the valency of the $i$-th vertex. Since we must connect all the legs sticking out of a vertex with a propagator, we have
\beq
\frac{\sum\limits_{i=1}^{n}{k_{i}}}{2}=p \Rightarrow \sum_{i=1}^{n}{k_{i}}=3n~,
\eeq{}
where we have used $\eqref{p3halvan}$. Since $k_i\geq 3$ $\forall i$, we draw the conclusion that $k_i=3$ $\forall i$.

Finally, there will be no tadpoles, since if we connect a vertex with itself we get identically zero since we are contracting two legs of a vertex, which are symmetric (resp. anti-symmetric) with the symplectic form, which is anti-symmetric (resp. symmetric).

In summary, the only vertices which will contribute will be 3-valent, they will be even in number, and there will be no tadpoles.
We have reproduced the counting argument of Rozansky and Witten but in a quicker way. For the readers who feel a certain qualm about the
viability of the action (\ref{RW_xi_c}), we hope we have demonstrated that the perturbation theory of (\ref{RW_xi_c})
  is exactly the same as the action (\ref{rwactionfinal}), which is a perfectly healthy physical theory.

\subsection{Perturbative Expansion of the BF-RW and CS-RW Model}

In fact a similar analysis can be applied to general 3D AKSZ models.  First we have to apply the exponential
 map in order to reduce problem to a linear space and then we can use the Hodge decomposition for the gauge
  fixing at the level of superfields. The counting argument about  trivalent graphs
   and the absence of tadpoles presented in previous subsection  are applicable to general 3D AKSZ models.
    Thus this gauge fixing scheme restricts all the Feynman diagrams to 3-valent graphs with no tadpoles. The statement may change however, when the source manifold has $H^1(\Sigma_3)\neq0$ and when external legs are allowed (the external legs are occupied by the harmonic modes). But in this paper we consider only $\Sigma_3=S^3$.

\begin{figure}[h]
\begin{center}
\includegraphics[bb=0 0 170 165,width=1in]{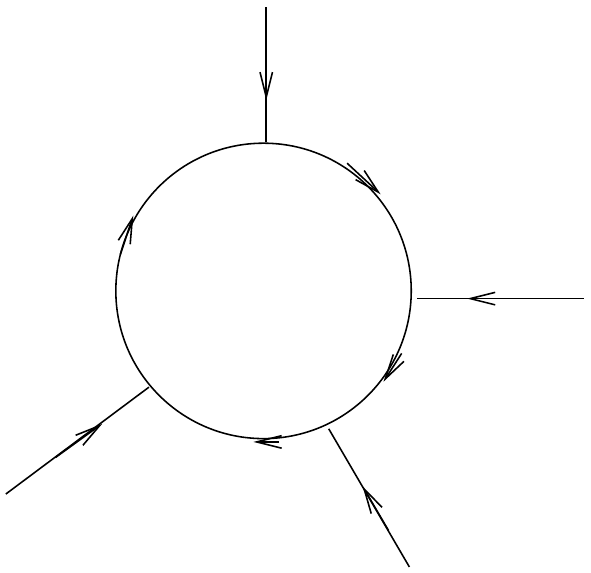}
\caption{The only diagram involving $BA$ propagator}\label{BA_sector_fig}
\end{center}
\end{figure}

 Since the analysis is straightforward we skip the detailed derivation of perturbative expansion for BF-RW and
  CS-RW models and give just a summary.
The perturbation expansion for the BF-RW model is rather simple. One sees from the kinetic term of the gauge sector that $\BS{A}$ can be connected to $\BS{B}$ and the only Feynman diagram involving $\BS{A,B}$ nontrivially is fig.\ref{BA_sector_fig}.
In this figure, an incoming arrow represents $\BS{A}$ and an outgoing one is $\BS{B}$. The external legs are occupied by $A^h_{(0)}$. Note that a harmonic zero form is but a constant. Hence technically this diagram is still two valent and thus vanishes. As a consequence, the $\BS{A}$ can only appear as $A^h_{(0)}$ tagged onto the 3-valent vertex from the term $\mu_A (A^A)^h$, in the same way that $(v^{\bar i})^h$ is attached to the curvature term.  Finally we end up
 with the following vertices
\beq
\left(R_{\bi j~l}^{~~n}\Omega_{nk} (\textsl{x}_0) v^{\bi}_{(0)}  + \nabla_{j}\nabla_{k}\partial_{l}\mu_{A}(\textsl{x}_0) A^{A}_{(0)}\right)\BS{\xi}^{j}\BS{\xi}^{k}\BS{\xi}^{l}~,
\eeq{slsl33e0d}
 where $v_{(0)}^{\bi}$ and $A^{A}_{(0)}$ are harmonic zero modes and we choose not to integrate over them.
 The $\BS{\xi}$'s will be connected by the propagator (\ref{propagator-xi}) according to the trivalent graphs
  without tadpoles. The resulting answer for partition function (without integration over zero modes)
   will be
  \beq
     \sum\limits_{\Gamma} b_\Gamma ~ c_\Gamma (\textsl{x}_0, v^{\bi}_{(0)}, A_{(0)}^A)~,
  \eeq{393osldk3}
    where $b_\Gamma$ is a kinematic factor (including the combinatorics)
      coming from the integration of the propagator (\ref{propagator-xi}) according
    to the trivalent  graph $\Gamma$ and they are obviously exactly the same as in RW model.
  The factor   $c_\Gamma (\textsl{x}_0, v^{\bi}_{(0)}, A_{(0)}^A)$ is made by  contracting the vertex (\ref{vertex-BF-CS})
     with $\Omega^{ij}$ according to the graph $\Gamma$.  The zero modes $A_{(0)}$ and $v^{\bar i}_{(0)}$ are both
     odd and thus the above perturbation expansion will terminate at some order.

Now let us discuss briefly the perturbative expansion for CS-RW model. In contrast with BF-RW model,
 the gauge field $\BS{A}$ participates non-trivially in the perturbation theory of the CS-RW model and the perturbative expansion continues to all orders. We will end up with the following vertices in CS-RW theory:
\bea
&&\frac16\left(R_{\bi j~l}^{~~n}\Omega_{nk} (\textsl{x}_0) v_{(0)}^{\bi} + \nabla_{j}\nabla_{k}\partial_{l}\mu_{A} (\textsl{x}_0)
A^{A}_{(0)}\right)\BS{\xi}^{j} \BS{\xi}^{k}\BS{\xi}^{l}~,\label{sskk33220} \\
&& \frac12\nabla_{i}\partial_{j}\mu_{A} (\textsl{x}_0) \BS{A}^{A}\BS{\xi}^{i} \BS{\xi}^{j}~,\label{33ik33qmd} \\
&& \frac{1}{6}f_{ABC}\BS{A}^{A}\BS{A}^{B}\BS{A}^{C}~\label{dkdkdkkkk23}~,
\eea
 where the superfields $\BS{A}$ and $\BS{\xi}$ are assumed to be co-exact and $v^{\bi}_{(0)}$, $A^A_{(0)}$,
  $\textsl{x}_0$ are constant zero modes.  Now in addition to the propagator (\ref{propagator-xi}) we have to introduce
   a super-propagator for $\BS{A}$
   \bea (\eta^{-1})^{AB}G(x_1,\theta_1;x_2,\theta_2)=\bra\BS{A}^A(x_1,\theta_1),\BS{A}^B(x_2,\theta_2)\ket~,\label{propagator-AA}\eea
  where $G$ is the same as in (\ref{propagator-xi}).
    Again we end up with the following perturbative expansion for the partition function (without the integration
     over zero modes)
       \beq
     \sum\limits_{\Gamma} b_\Gamma ~ c_\Gamma (\textsl{x}_0, v^{\bi}_{(0)}, A_{(0)}^A)~,
  \eeq{393osldk3ss}
    where $c_\Gamma$ are constructed by contracting  three vertices
      presented above by $\eta^{AB}$ and $\Omega^{ij}$
     according to the trivalent graph $\Gamma$. Now we have to look at all possible $\Gamma$ with different vertices
      (\ref{sskk33220})-(\ref{dkdkdkkkk23}) and the sum (\ref{393osldk3ss}) contains all orders now. However,
       the vertex   (\ref{sskk33220}) will enter the diagrams in very restrictive fashion since the zero modes $v_{(0)}^{\bi}$
        and $A_{(0)}^A$ are odd.
     Later we will give concrete example of $c_\Gamma$.

\section{AKSZ Models and Characteristic Classes}
\label{char-classes}

We have come to the second part of the paper, where we shall discuss the mathematical aspects of our models, in
 particular the interpretation of the perturbative partition function.
 To be reasonably self contained, we shall review quickly the heuristic ideas of our approach developed in previous
  work. For more detailed discussion the reader may consult  \cite{Qiu:2009rf}.

To construct characteristic classes for flat bundles, it is very convenient to use the Chern-Weil homomorphism
applied to flat connections \cite{BottHaefliger,Fuks}. Let ${\mathbf g}^*$ be the dual of a Lie algebra for some Lie group
 ${\mathbf G}$. One can form the
Chevalley-Eilenberg (CE) complex $(\wedge^{\cdot} {\mathbf g}^*,\delta)$ as follows:  supposing $c^q\in\wedge^{q+1}{\mathbf g}^*$ and $A_0,A_1,\cdots$ are elements of the Lie algebra basis,
then $\delta c^q$, when evaluated on $A_0,A_1,\cdots$, can be written as
\bea \delta c^q (A_0,\cdots,A_q,A_{q+1})=\sum_{i<j}(-1)^{i+j+1}c^q([A_i,A_j],A_0,\cdots \hat{A_i},\cdots \hat{A_j},\cdots A_{q+1})~.\nn\eea
If one is given a flat connection $A$ for some principle bundle
(meaning $A$ is some Lie algebra valued one form satisfying $dA+\frac{1}{2}[A,A]=0$), one can evaluate the cohain $c^q$ on the connection and obtain a $(q+1)$-form. The de Rham differential acting on such a form gives
\bea d c^q (\underbrace{A,\cdots A}_{q+1})=-\frac{q+1}{2}c^q([A,A],\underbrace{A,\cdots A}_{q-1})
=-\frac{1}{q+2}\delta c^q (\underbrace{A,\cdots A}_{q+2})~.\nn\eea
In other words, if $c^q$ is $\delta$-closed, the resulting form by plugging in $A$ is closed, and further it can be shown that if one changes the flat connection, the form varies by an exact form. This is the Chern-Weil homomorphism.

In the super language, the de Rham operator is written as $v^{\mu}\partial_{\mu}$, and can be regarded as a vector field on the graded manifold $T[1]M$ (where $v^{\mu}$ serves as the degree 1 coordinate of the fibre of $TM$). In general, $d$ can be replaced by any nilpotent vector field $Q$ on a graded manifold, and the flatness condition can be replaced by an appropriate Maurer-Cartan equation: $QA+1/2[A,A]=0$. We can then repeat everything in the construction above and obtain classes that are $Q$-closed, the $Q$-characteristic classes.

The AKSZ models contain both ingredients of the above construction: a cocycle and a flat connection.
Using the notations for a general AKSZ model from subsection \ref{Review-RW}, we now briefly review the idea.
First, the interaction part of the action, which we now call $\Theta$, is an odd function on some even graded symplectic manifold $({\cal M},\omega)$. The master equation implies $\{\Theta,\Theta\}=0$, implying that the Hamiltonian vector field generated by $\Theta$ is a candidate for $Q$, $Q^B=(\Theta\overleftarrow{\partial}_A)(\omega^{-1})^{AB}$. For simplicity, we assume that the degrees are all carried by the coordinates $\Phi$ and $\omega$ is a constant of even degree. Moreover we assume that ${\cal M}$ is a vector space (if this is not the case, we have to
 apply the exponential map).
Following \cite{Qiu:2009rf}, we expand the equation
\bea(\Theta\overleftarrow{\partial}_A) (\omega^{-1})^{AB}\partial_B\Theta=0\nn\eea
into a power series around a fixed point $\Phi^A=\Phi_0^A+\phi^A$. By ordering the $\phi$'s wisely, we have a neat expression without sign factors,
{\small\bea &&~~~2(\phi^{C_1}\partial_{C_1}\Theta\overleftarrow{\partial}_A)(\omega^{-1})^{AB}\partial_B\Theta\nn\\
&&+\big(\phi^{C_2}\phi^{C_1}
(\partial_{C_1}\partial_{C_2}\Theta\overleftarrow{\partial}_A)(\omega^{-1})^{AB}\partial_B\Theta+\phi^{C_2}
(\partial_{C_2}\Theta\overleftarrow{\partial}_A)(\omega^{-1})^{AB}\phi^{C_1}\partial_{C_1}\partial_B\Theta\big)+\cdots\nn\\
&&+\sum_{p=0}^n\frac{1}{p!(n-p)!}\phi^{C_p}\cdots\phi^{C_1}
(\partial_{C_1}\cdots\partial_{C_p}\Theta\overleftarrow{\partial}_A)(\omega^{-1})^{AB}\phi^{C_n}\cdots\phi^{C_{p+1}}
(\partial_{C_{p+1}}\cdots\partial_{C_n}\partial_B\Theta)\nn\\
&&+\cdots,\nn\eea}
where all the derivatives are evaluated at the point $\Phi_0$.
The coefficient of each power of $\phi$ has to vanish and we get a series of equations.
By defining some short-hands
\bea \Theta_m=\phi^{C_m}\cdots\phi^{C_1}\frac{1}{m!}\partial_{C_1}\cdots\partial_{C_m}\Theta(\Phi_0)~,~~~ \Theta^m=\sum_{n=m}^{\infty}\Theta_n~,\label{short_hand}\eea
we can rewrite the equations as
\bea &&\{\Theta_1,\Theta_2\}=0,~~ \{\Theta_1,\Theta^3\}+\frac{1}{2}\{\Theta^2,\Theta^2\}=0~,\nn\eea
where $\{~,~\}$ are Poisson brackets in $\phi$-variable.
The first equation says $Q^A\partial_AQ^B=0$, \ie $Q^2=0$. The second is best rewritten as
\bea Q^A\frac{\partial}{\partial\Phi_0^A}\Theta^2+\frac12\{\Theta^2,\Theta^2\}=0~\label{Maure_Cartan}.\eea
This is essentially because, for example, at order 3
\bea \{\Theta_1,\Theta_3\}=\Theta\overleftarrow{\partial}_B(\omega^{-1})^{BA}\frac12\phi^C\phi^D\partial_D\partial_C\partial_A\Theta
=Q^A\frac12\phi^C\phi^D\partial_D\partial_C\partial_A\Theta=Q^A\frac{\partial}{\partial\Phi_0^A}\Theta_2~.\label{globalization}\eea
This step is true for a flat target manifold, since the ordinary derivatives (graded) commute. For the curved case, the power series expansion
is replaced with an exponential map that identifies a neighborhood of $\Phi_0$ with the tangent space at $\Phi_0$. As a result the last rewriting must be checked by hand. For the BF-RW and CS-RW models on a hyperK\"ahler manifold, however, (\ref{globalization})
 stand correct. We will discuss more about this issue of globalization in section \ref{math}, where we relax the hyperK\"ahler condition.

Back from our digression, (\ref{Maure_Cartan}) is just the Maurer-Cartan equation and $\Theta^2$ is a flat connection on $(\Phi_0, \phi) \in {\cal M} \times {\cal M}$
 where in the $\phi$-directions we have the bracket $\{\cdot,\cdot\}$ and the $\Phi_0$-directions we treat as a base.

The second ingredient for the Chern-Weil homomorphism is a cocycle. That will be provided by the path integral, more
 precisely by its perturbative expansion which is well defined.

We look at an AKSZ theory with only the kinetic term
\bea \int_{{\cal L}}~D\BS{\Phi}~\exp\Big(\smalint d^6z~\BS{\Phi}^A\omega_{AB}D\BS{\Phi}^B\Big)~.\nn\eea
Here we assume that the $\Phi$ takes value in a flat symplectic manifold $({\cal M},\omega)$
(otherwise we use the exponential map for the expansion).
For a collection of functions $f_0,\cdots f_q$ on ${\cal M}$, consider the correlator,
{\small\bea c^q(f_0,f_1,\cdots f_q)=
\int_{{\cal L}}~D\BS{\Phi}~\smalint d^6z_0f_0(\BS{\Phi})~\smalint d^6z_1f_1(\BS{\Phi})\cdots \smalint d^6z_qf_q(\BS{\Phi})~\exp\Big(\smalint d^6z~\BS{\Phi}^A\Omega_{AB}D\BS{\Phi}^B\Big).\label{cochain}\eea}
This correlator can be thought of as a cochain in the CE complex of the Lie algebra of Hamiltonian vector fields on ${\cal M}$. By using the property that the
integral of some $\Delta$-exact function over a Lagrangian submanifold is zero, and also the properties (\ref{BV_bracket}), (\ref{property_Laplacian})
we have
\bea 0&=&\int_{{\cal L}}~D\BS{\Phi}~\Delta\Big[\smalint d^6z_0f_0(\BS{\Phi})~\smalint d^6z_1f_1(\BS{\Phi})\cdots \smalint d^6z_qf_q(\BS{\Phi})~\exp\big(\smalint d^6z~\BS{\Phi}^A\Omega_{AB}D\BS{\Phi}^B\big)\Big]\nn\\
&=&\sum_{i<j}(-1)^{f_i+s_{ij}}c^{q-1}(\{f_i,f_j\},f_0,\cdots,\hat{f_i},\cdots \hat{f_j},\cdots f_q)\nn\\
&=&\delta c^q(f_0,\cdots,f_q)~,\label{codiff}\eea
where we abbreviate $(-1)^{\deg\,f_i}$ as $(-1)^{f_i}$ and define the Kozul sign
\bea s_{ij}=(f_i+1)(f_0+\cdots f_{i-1}+i)+(f_j+1)(f_0+\cdots f_{j-1}+j)+(f_i+1)(f_j+1)~.\nn\eea
The details of the derivation can be found in \cite{Qiu:2009rf}. In conclusion,
the cochain (\ref{cochain}) provided by the path integral is a cocycle. It can be shown that if one changes the choice of ${\cal L}$,
one only changes the cocycle by a coboundary.

This cocycle condition is derived when we perform the path integral over all the fields, but in practice,
the integration of zero modes can be tricky. One can instead refrain from integrating over zero modes,
leaving them as parameters. This way the cochains no longer take value in the real numbers but rather become cochains of the Lie algebra of Hamiltonian vector fields taking values in $C^{\infty}({\cal M})$, and thereby giving a richer structure. In such cases, the above cocycle condition is modified. Here we only write down the differential operator acting on a 1-cocycle for later use
\bea \delta c^1(f,g,h)&=&(-1)^fc^1(\{f,g\},h)-(-1)^{f\deg c^1}\{f,c^1(g,h)\}\nn\\
&&+(-1)^{f+(h+1)(g+1)}c^1(\{f,h\},g)-(-1)^{(g+1)(f+1)+g\deg c^1}\{g,c^1(f,h)\}\nn\\
&&+(-1)^{g+(f+1)(g+h)}c^1(\{g,h\},f)-(-1)^{(h+1)(f+g)+h\deg c^1}\{h,c^1(f,g)\}~.\label{codiff_mod}\eea
In this expression, $\deg c^1$ is the degree of $c^1$, namely $\deg c^1(f,g)=\deg\, f+\deg\, g+\deg c^1$.
For the cochains of type (\ref{cochain}), the degree is zero.

Now we have all the ingredients in the Chern-Weil homomorphism, a cocycle and a flat connection, and we can consider the perturbative expansion of the model defined by
\bea
S=\int d^6z ~ \Big(\frac12\BS{\Phi}^A\omega_{AB}D\BS{\Phi}^B+\BS{\Theta}\Big)~.\label{generic_AKSZ}
\eea
At each order of perturbation, we have a correlator
\bea
Z (\Phi_0)=\sum_{q=1}^{\infty}\hbar ^q c^{2q-1}(\Theta^2,\cdots,\Theta^2)~,\nn
\eea
where $\Theta^2$ is defined in (\ref{short_hand}). Each term $c^{2q-1}$ in the expansion is $Q$-closed,
\bea
Qc^{2q-1}(\Theta^2,\cdots,\Theta^2)=-qc^{2q-1}(\{\Theta^2,\Theta^2\},\cdots,\Theta^2)
=\frac{1}{2(2q-1)}\delta c^{2q-1}(\Theta,\cdots \Theta)=0\nn
\eea
and furthermore, the change in ${\cal L}$ only effects a change in $Q$-exact terms.
We conclude that $c^{2q-1}(\Theta,\cdots \Theta)\in H_Q^{2q}(M)$. The class represented by $c^{2q-1}$ is independent of the choice ${\cal L}$ (note that all the extraneous data such as the metric of the target or source manifold comes in only through ${\cal L}$). The above discussion applies to a general AKSZ model.

Moreover the partition function can be written as a sum over graphs
\beq
  Z(\Phi_0) = \sum\limits_{\Gamma} b_\Gamma c_\Gamma(\Phi_0)~,
\eeq{sslsl330dirr}
 where  $b_\Gamma$ is given by integrals over correlators and $c_\Gamma (\Phi_0)$ is constructed by
  contracting $\partial_{C_1}\cdots\partial_{C_m}\Theta(\Phi_0)$ by $(\omega^{-1})^{AB}$ according to the graph $\Gamma$.
   It is important that
   \beq
    \sum\limits_{\Gamma} c_\Gamma (\Phi_0) \Gamma
   \eeq{ssl3333}
 is a graph cycle up to $Q$-exact terms. In fact the data $(\Theta, \omega)$ on ${\cal M}$
   constitutes an $L_{\infty}$ algebra structure and for such a structure one can always construct graph cycles \cite{Kontsevich:noncommutative}. If we restrict ourselves only to trivalent graphs (this is what our perturbative theory naturally produces),
  then this statement says that
  $c_\Gamma(\Phi_0)$ satisfies IHX-relation up to $Q$-exact terms. Consequently, the partition function and therefore the characteristic classes
  (\ref{sslsl330dirr}) only depend on the graph cohomology class of $\sum_{\Gamma} b_{\Gamma}\Gamma^*$. More details about the interplay between $b_{\Gamma}$ and $c_{\Gamma}$ can be found in \cite{Qiu:2009rf}.

\subsection{Characteristic Classes from the RW, BF-RW and CS-RW Models}

For the RW model, recall that we have the MC equation (\ref{key_id}), with $Q_{RW}=\bar\partial$ and the RW classes
are elements of the Dolbeault cohomology. These class are related to the so called holomorphic foliation. For such
cases, we have a principle bundle structure, the gauge group is formal holomorphic symplectomorphisms and its Lie algebra is formal holomorphic Hamiltonian vector fields. Here formal means that the Hamiltonian function that generates these vector fields are formal power series in $\xi^i$. It also turns out that our function $\Theta_{\bar i}(\xi)$ is the flat connection arising from the foliation problem as prescribed by Fuks in \cite{Fuks,FuksI}, the algorithm is clearly laid out in the latter reference. The same procedure is come upon independently by a number of authors \cite{BernsteinRosenfeld,BottHaefliger}, in \cite{BottHaefliger} it is also explained the need to restrict the cocycle to be basic, that is, the cocyle is invariant under $sp(2n)$ transformations, and if any of the functions $f_i$ in (\ref{cochain}) is quadratic, the cochain vanishes. Both properties are clearly satisfied by our path integral construction and are crucial for the later analysis of invariance properties of our characteristic classes. This interpretation of the RW invariants was outlined in \cite{Kontsevichformal}, further
clarified in \cite{kapranov-1997,Qiu:2009rf}.

\subsection{BF-RW Model}

We can slightly generalize the previous construction to incorporate a gauge field. For the RW model on a hyperK\"ahler manifold coupled to BF theory, defined in
(\ref{BFRWaction}), we would like to apply the same exponential map (\ref{exp_map}). The resulting action is
\bea
&&S=\int d^{6}z~\Big(\BS{B}_{A}D\BS{A}^{A}+ \BS{p}_{\bi}D\BS{X}^{\bi}+\BS{q}_{\bi}D \BS{v}^{\bi}\nn\\
&&\hspace{2.3cm}+\frac{1}{2}\Omega_{ij}\BS{\xi}^{i}D\BS{\xi}^{j}+\frac{1}{2}f_{AB}^{~~~C}\BS{A}^{A}\BS{A}^{B}\BS{B}_{C}-\BS{p}_{\bi}\BS{v}^{\bi}
+\BS{v}^{\bar i}\BS{\Theta}_{\bar i}+\mathsf{M}^0_A\BS{A}^{A}\big)~,\nn\eea
where $\Theta_{\bi}$ is defined in (\ref{def_Theta}) and
\bea\mathsf{M}^n_A=\sum_{k=n}^{\infty}\SF{M}_{Ak},~~\SF{M}_{Ak}=\frac{1}{k!}\xi^{i_1}\cdots \xi^{i_k}\nabla_{i_k}\cdots \nabla_{i_2}\partial_{i_1}\mu_A(\textsl{x}_0)~,\label{def_M_A}\eea
 where we use the hyperK\"ahler properties.
Some of its properties are
\bea \bar\partial\mathsf{M}^2_A=\bar\partial\mathsf{M}^1_A=\bar\partial\mathsf{M}^0_A=-\{v^{\bar i}\Theta_{\bar i},\mathsf{M}^1_A\}~;
~~~\{A^A\mathsf{M}^1_A,A^B\mathsf{M}^1_B\}=A^AA^Bf_{AB}^{~~C}\mathsf{M}^0_C~.\label{prop_mu}\eea

We need to derive a Maurer-Cartan equation in this case. One can of course follow the procedure as for (\ref{Maure_Cartan}),
but we would like to also treat the gauge fields as parameters before choosing a gauge fixing.
We need to modify somewhat the procedure that leads to (\ref{Maure_Cartan}). To this end,
assume that the symplectic form of the target space splits into two independent parts $\omega=\omega_p+\Omega$,
where $\omega_p$ is for the 'parameter sector' and $\Omega$ is for the fields that are active in perturbation (\ie only $\xi$).
Again expanding the equation
\bea
\{\Theta,\Theta\}_{\omega_p}+\{\Theta,\Theta\}_{\Omega}=0~,\nn\eea
into a power series of the \emph{active fields only}, we obtain the equation series
\bea
&&\{\Theta^0,\Theta^0\}^1_{\omega_p}+2\{\Theta_1,\Theta^2\}_{\Omega}+\{\Theta^2,\Theta^2\}_{\Omega}=0~,\nn\\
&&\{\Theta^0,\Theta^0\}^2_{\omega_p}+2\{\Theta_1,\Theta^3\}_{\Omega}+\{\Theta^2,\Theta^2\}_{\Omega}=0~,\nn\\
&&\{\Theta^0,\Theta^0\}^3_{\omega_p}+2\{\Theta_1,\Theta^4\}_{\Omega}+2\{\Theta_2,\Theta^3\}_{\Omega}+\{\Theta^3,\Theta^3\}_{\Omega}=0~.\label{MC_w_para}\eea
Pay attention that the numbers in the super or subscript only denote the powers of the active fields, not the parameters. This is the Maurer-Cartan
equation modified due to our needs to treat certain fields as parameters. We have chosen to go up to fourth order, because we know that eventually the vertices appearing in the perturbation will be of cubic order.

For the BF-RW model, we treat $p_{\bi},X^{\bi},q_{\bar i},v^{\bar i},A,B$ as parameters, while $\xi^i$ alone is active. Further notice
\footnote{Due to unfortunate notation, there is a likely confusion between $\Theta$, which is a generic interaction term for the
action (\ref{generic_AKSZ}), and $\Theta_{\bar i}$ defined in (\ref{def_Theta}).}
{\small\bea \{\Theta^0,\Theta^0\}^3_{\omega_p}=2\bar\partial(v^{\bar i}\Theta_{\bar i}
+A^A\mathsf{M}^3_A)-A^AA^Bf^{~~C}_{AB}\mathsf{M}^3_C=2(\bar\partial-\frac12A^AA^Bf^{~~C}_{AB}\frac{\partial}{\partial A^C})
\big(v^{\bar i}\Theta_{\bar i}+A^A\mathsf{M}^3_A\big)~,\nn\\
\{\Theta_1,\Theta^4\}_{\Omega}=\{A^A(\partial_i\mu_A)\xi^i,v^{\bar i}\Theta^4_{\bar i}+A^A\mathsf{M}_A^4\}=
A^Ak_A^i\nabla_{\textsl{x}^i_0}\big(v^{\bar i}\Theta_{\bar i}+A^A\mathsf{M}_A^3\big)~,\nn\eea}
and we obtain our modified Maurer-Cartan equation
\bea
&&Q\big(v^{\bar i}\Theta_{\bar i}+A^A\mathsf{M}_A^3\big)+\frac{1}{2}\big\{v^{\bar i}\Theta_{\bar i}+A^A\mathsf{M}_A^3,v^{\bar i}\Theta_{\bar i}+A^A\mathsf{M}_A^3\big\}+\big\{A^A\mathsf{M}_{A2},v^{\bar i}\Theta_{\bar i}+A^A\mathsf{M}_A^3\big\}=0~,\nn\\
&&Q=(\bar\partial+A^Ak_A^i\nabla_{\textsl{x}^i_0}-\frac12A^AA^Bf^{~~C}_{AB}\frac{\partial}{\partial A^C})~,\label{Q_BF}\eea
 where the brackets $\{~,~\}$ are taken in $\xi$-direction and in $Q$ the covariant derivative $\nabla_{\textsl{x}^i_0}$ can be replaced
  by the partial derivative when it acts on $(0,p)$-forms.
So the MC equation is satisfied up to a $sp(2n)$ rotation in the $\xi^i$ space (the last term of the first line). Yet from our earlier discussion, the cochain defined
by the path integral is invariant with respect  to $sp(2n)$ rotations. Also, $Q$ does not square to zero but rather a curvature term. This is again not a problem, since for our purpose, $Q$ will always act on $(0,p)$ forms while the connection is of pure index. Finally, if one does not trust the above derivation, by using the property of $\mathsf{M}$ in (\ref{prop_mu})
one can check that the MC equation is indeed fulfilled.

Proceeding to the lowest order of the perturbative expansion, the Feynman rules gives for the following graph
\begin{figure}[h]
\begin{center}
\includegraphics[bb=0 0 113 51,width=.8in]{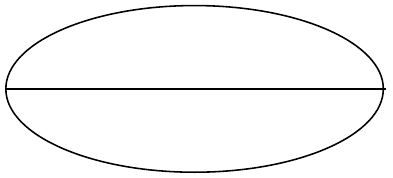}
\caption{$\Gamma$, perturbation expansion to the lowest order}\label{lowestorder_fig}
\end{center}
\end{figure}
\bea c^1(\Theta,\Theta)=\frac16(\Theta\overleftarrow{\partial}_{A_1}\overleftarrow{\partial}_{A_2}\overleftarrow{\partial}_{A_3})(\Omega^{-1})^{A_1B_1}
(\Omega^{-1})^{A_2B_2}(\Omega^{-1})^{A_3B_3}(\partial_{B_1}\partial_{B_2}\partial_{B_3}\Theta).\label{lowestorder}
\eea
If we specialize to our BF-RW theory, the interaction term $\Theta=v^{\bar i}\Theta_{\bar i}+A^A\mathsf{M}^3_A$ plays the role of the 3-valent vertices and we obtain
{\small\bea c_{\Gamma}(\textsl{x}_0,v,A)=\frac{1}{6}\partial_i\partial_j\partial_k(v^{\bar i}\Theta_{\bar i}+A^A\mathsf{M}^3_A\big)(\Omega^{-1})^{jo}(\Omega^{-1})^{kp}(\Omega^{-1})^{lq}
\partial_o\partial_p\partial_q\big(v^{\bar i}\Theta_{\bar i}+A^A\mathsf{M}^3_A\big)\nn\\
=\frac16\big(R^{~~n}_{\bi j~l}v^{\bi}\Omega_{nk}+A^{A} \nabla_{j}\nabla_{k}\partial_{l}\mu_{A}\big)
(\Omega^{-1})^{jo}(\Omega^{-1})^{kp}(\Omega^{-1})^{lq}
\big(R_{\bi o~q}^{~~r}v^{\bi}\Omega_{rp}+ A^A\nabla_{o}\nabla_{p}\partial_{q}\mu_{A}\big)~.\label{ch_class_BFRW}\eea}
From the general arguments given in the beginning of section \ref{char-classes}, we have $Qc_{\Gamma}=0$. Indeed one can check explicitly that $Qc_{\Gamma}=0$ using the above concrete expressions for $Q$ and $c_{\Gamma}$. This is a straightforward but lengthy
  calculation. $c_{\Gamma}$ is merely a representative of the $Q$-cohomology class when one uses the
 hyperK\"ahler metric.
When using a general metric, the expression for $c_{\Gamma}$ will be much more complicated, but still lies in the same $Q$-class.

\subsection{CS-RW Model}

In this case, the gauge fields $\BS{A}$ does take part in the perturbation theory actively, since now the propagators in the gauge sector is $\bra\BS{A}(x,\theta),\BS{A}(y,\xi)\ket$ in contrast to $\bra\BS{A}(x,\theta),\BS{B}(y,\xi)\ket$ in BF theory. The application of the exponential map is exactly the same as before, and we follow the same procedure as in the previous section to derive the Maurer-Cartan equation. The gauge field is now split as $\BS{A}=\BS{A}^h+\BS{A}^c$, the superscripts ${}^h,{}^c$ suggest harmonic and coexact.
   Let us abuse the notation a little and use $A^h$ as base coordinate and $A^c$ as fiber coordinate in the Taylor expansion
    (see section \ref{char-classes} for the expansion into a power series around a fixed point $\Phi^A=\Phi_0^A+\phi^A$).
Note that the expansion is slightly different
\bea &&\Theta^2=(A^A)^h\mathsf{M}_A^2+(A^A)^c\mathsf{M}_A^1+\frac12f_{ABC}(A^A)^h(A^B)^c(A^C)^c,\nn\\
&&\Theta^3=v^{\bar i}\Theta_{\bar i}+(A^A)^h\mathsf{M}_A^3+(A^A)^c\mathsf{M}_A^2+\frac16f_{ABC}(A^A)^c(A^B)^c(A^C)^c~.\nn\eea
The Maurer-Cartan equation reads
\bea &&\bar\partial\Theta^3+\big(-\frac12f_{ABC}(A^A)^h(A^B)^h\frac{\partial}{\partial (A^C)^h}-\mu_A\eta^{BA}\frac{\partial}{\partial (A^A)^h}+(A^A)^hk_A^i\nabla_i\big)\Theta^3\nn\\
&&+\{\Theta_2,\Theta^3\}+\frac12\{\Theta^3,\Theta^3\}=0~,\label{Maurer_Cartan_CS}\eea
where the bracket $\{~,~\}$ is understood in $\xi$- and $A$-directions.
There are two features to this equation. Firstly it is no longer homogeneous in powers of $A$ and $v$; secondly the next to last term now represents an $osp(2n|m)$ rotation where $2n$ is the complex dimension of the manifold and $m$ is the rank of the gauge group.

For the graph fig.\ref{lowestorder_fig}, the Feyman rules give (we have dropped the superscript ${}^h$ on $A$, for it is clear that the $A$'s that are uncontracted must be the harmonic modes)
{\small\bea &&c_{\Gamma}(\textsl{x}_0,v,A)\nn\\
&=&\frac16\big(R_{\bi j~l}^{~~n}v^{\bi}\Omega_{nk}+A^{A} \nabla_{j}\nabla_{k}\partial_{l}\mu_{A}\big)
(\Omega^{-1})^{jo}(\Omega^{-1})^{kp}(\Omega^{-1})^{lq}
\big(R_{\bi o~q}^{~~r}v^{\bi}\Omega_{rp}+ A^A\nabla_{o}\nabla_{p}\partial_{q}\mu_{A}\big)\nn\\
&&+\frac12\nabla_{j}\partial_{k}\mu_{A}\nabla_{m}\partial_{n}\mu_{B}\Omega^{jm}\Omega^{kn}\eta^{AB}~.\nn\eea}
Again, by general arguments this expression should $Q$-closed, where $Q$ is given by
\bea
Q=\bar\partial-\frac12f_{ABC}A^AA^B\frac{\partial}{\partial A^C}+\mu_A\eta^{BA}\frac{\partial}{\partial A^A}+A^Ak_A^i\nabla_i~,\label{Q_CS}
\eea
where the covariant derivative can be replaced by ordinary ones when it acts on $(0,p)$-forms. Again, it is a straighforward but lenghty calculation to check this explicitly.

\subsection{Some Explicit Checks}

 Now let us return to a discussion of general 3D AKSZ theories.
  We can argue that any trivalent graph gives rise to a cocycle of Hamiltonian vector fields.
First, let us again perform some quick checks on the statement $Qc_{\Gamma}=0$ from another perspective, and see how it
 can be generalized for any trivalent graph. As mentioned above, the Maurer-Cartan equations (\ref{Maure_Cartan}),  (\ref{Maurer_Cartan_CS}) are rather easy to check by using the properties (\ref{key_id}), (\ref{prop_mu}). What remains to be checked stands thus
\bea 0\stackrel{?}{=}Q c^1 (\Theta^3,\Theta^3)=-2c^1(\{\Theta_2,\Theta^3\},\Theta^3)-2c^1(\{\Theta^3,\Theta^3\},\Theta^3)~.\nn\eea
The first term on the right hand side is zero due to the $sp(2n)$ or $osp(2n|m)$ invariance of the cochain $c^1$. For the second term, by using the definition of the differential operator (\ref{codiff_mod}), it can be written as
\bea \delta c^1(\Theta^3,\Theta^3,\Theta^3)=-3c^1(\{\Theta^3,\Theta^3\},\Theta^3)-3\{\Theta^3,c^1(\Theta,\Theta)\}=
-3c^1(\{\Theta^3,\Theta^3\},\Theta^3)~.\nn\eea
The last term drops because the operation $\{\Theta^3,\cdot\}$ requires $\Theta^3$ to have non-vanishing first derivative in $\xi$ at $\xi=0$, but $\Theta^3$ starts off at order $\xi^3$.

So we only need to check whether the cochain $c^1$ is a cocycle. In the following, we explicitly check the case where all fields involved are purely bosonic, yet enough detail is given so that the reader can check the general case.
The cochain $c^1$, when evaluated on two functions, gives
\bea c^1(f,g)=(\partial_A\partial_B\partial_Cf)(\omega^{-1})^{AD}(\omega^{-1})^{BE}(\omega^{-1})^{CF}(\partial_D\partial_E\partial_Fg)~,\nn\eea
and the cocycle condition reads
\bea c^1(\{f,g\},h)-\{f,c^1(g,h)\}+\textrm{cyc perm}=0~,\nn\eea
where for the sake of clarity we assume that
 $f,g,h$ and the coordinates $\Phi^A$ are purely bosonic. For the general graded case the argument can be repeated.

Even for this simple formula, it is not recommended to check it by brute force. We can use a graphic method to simplify things. We denote $(\partial_Af)(\omega^{-1})^{AB}(\partial_Bg)$ as $f\to g$; $(\partial_A\partial_Bf)(\omega^{-1})^{AC}(\omega^{-1})^{BD}(\partial_C\partial_Dg)$ as
$f\Rightarrow g$, etc.  Reversing an arrow means a minus sign.
The cocycle condition amounts to
\bea \Big((f\to g)\Rrightarrow h\Big)-\Big(f\to(g \Rrightarrow h)\Big) + \textrm{cyc perm}=0~.\nn\eea
Then easily we get (the over-arrow is meant to go from $f$ to $h$)
\bea &&\Big((f\to g)\Rrightarrow h\Big)-\Big(f\to(g \Rrightarrow h)\Big)\nn\\
&=&\big(g\leftarrow f\Rrightarrow h\big)+\big(f\rightarrow g\Rrightarrow h\big)
-\big(f\rightarrow g\Rrightarrow h\big)-\big(\overrightarrow{f~~~g\Rrightarrow h}\big)\nn\\
&&+3\big(\overrightarrow{f\to g \Rightarrow h}\big)+3\big(\overrightarrow{f\Rightarrow h \leftarrow g}\big)~,\nn\eea
each of the terms on the right hand side will find their companion with a minus sign from the cyclic permutations.

In fact, using this graph method, it is not difficult to show that if a cochain is represented solely by 3-valent graphs, then it is closed. This result probably is already known to Kontsevich \cite{Kontsevich:noncommutative}. We stress that this does not trivialize our result, for what is more interesting than the closedness of these cochains is how they change under a change of gauge condition and metric etc. These properties follows non-trivially from the BV manipulations.

\section{Mathematical Comments}
\label{math}

In this section we collect some mathematical comments on the RW, BF-RW and CS-RW models. In particular we would like to
 explain briefly the general case of a holomorphic symplectic manifold. We present the construction of the corresponding characteristic classes
  and argue their independence from the connection used in the construction.

In the earlier subsection \ref{Review-RW},  we took the AKSZ action
\bea
S_{RW}=\int{d^{6}z~\left(\BS{p}_{\bar i}D\BS{X}^{\bar i}+\BS{q}_{\bar i}D \BS{v}^{\bar i}+\frac{1}{2}\Omega_{ij}\BS{X}^{i}D\BS{X}^{j}-\BS{p}_{\bar i}\BS{v}^{\bar i}\right)}~,\nn\eea
whose definition only depended on the complex structure and holomorphic symplectic form, and we specialized
to the case of a hyperK\"ahler manifold and obtained the RW classes.

Since the metric of the hyperK\"ahler manifold does not enter the defining data of the action, but rather comes in later through the exponential map (\ie through the choice of gauge fixing), by a general physical argument, we know that the RW classes do not depend on the
 hyperK\"ahler metric. In contrast, in the work by Kapranov \cite{kapranov-1997} and reviewed by Sawon \cite{sawon-phd},  the authors chose to integrate the RW class on the hyperK\"ahler manifold and obtain the RW invariants. After performing the integral, a somewhat stronger invariance property can be claimed. The key observation is that the 3-valent vertex in the perturbation theory is identified as the Atiyah class, which is the obstruction to the existence of a holomorphic connection. The Dolbeault representative of this class is just the (1,1) component of the curvature tensor. But for a holomorphic vector bundle with Hermitian metric one can always construct a connection such that the curvature is automatically (1,1). Thus the Atiyah class can be defined without reference to the complex structure. Furthermore, this class is by construction independent of the K\"ahler form $\omega$, and hence independent of $\textrm{Re}\Omega$ and $\textrm{Im}\Omega$ by the total democracy between $\omega,\textrm{Re}\Omega,\textrm{Im}\Omega$. This last statement leads to the claim that the RW invariants are constant
   on the connected component of the moduli space of hyperK\"ahler metrics. But from our approach, without integrating the RW class, the last claim can not be made.

On the other hand, we will now relax the hyperK\"ahler condition. From the earlier discussion of perturbation theory of the RW model, we saw a few salient features. Firstly, we needed to expand around a fixed point $\textsl{x}_0$ using the exponential map. It was crucial that there the Levi-Civita connection preserves $\Omega$, the symplectic form $\Omega_{ij}(X)\delta X^i\delta X^i$ is pulled back neatly into $\Omega_{ij}(\textsl{x}_0)\delta{\xi}^i\delta{\xi}^i+2\delta\Theta_{\bar i}\delta X^{\bar i}$, \ie after a shift of $p_{\bar i}$ that absorbs the second term, the $\xi$ fields live in $\mathbb{C}^{2n}$ with a constant symplectic form. Thus our simple gauge fixing condition $\BS{\xi}^e=0$ can be applied. Secondly, to obtain the relevant Maurer-Cartan equation, we pulled back the $\bar\partial$ and showed that it is given by $\bar\partial+v^{\bar i}\{\Theta_{\bar i},\cdot\}$ in the pull back frame. Once we relax the hyperK\"ahler condition, we still would like to preserve these features, for otherwise, the structure of perturbation theory changes drastically and we can no longer make reliable claims about metric independence.

For an integrable complex structure, it is known that one can construct a torsionless connection, such that $\nabla J=0$ (e.g., see \cite{MR794310}). Such a connection, written in complex coordinates, has the only non-zero components $\Gamma^i_{\mu j}$ or $\Gamma^{\bar i}_{\mu \bar j}$. Then by the torsionless condition $\Gamma_{\bar i l}^j=\Gamma_{l\bar i}^j=0$, \ie $\Gamma$ is of purely holomorphic or anti-holomorphic indices. Based on this connection, we can further construct another one such that $\Omega$ is preserved. Let
\bea
\tilde\Gamma_{ij}^k=\Gamma_{ij}^k+\frac13\Omega^{kl}\big(\nabla_i\Omega_{lj}+\nabla_j\Omega_{li}\big)~.\nn
\eea
The new connection has the same transformation property as the original one, since the additional piece is covariant. It is torsionless and preserves $\Omega$
\bea \tilde\nabla_i\Omega_{jk}&=&\nabla_i\Omega_{jk}-\frac13\Omega^{lm}\big(\nabla_i\Omega_{m[j}+\nabla_{[j}\Omega_{|mi}\big)\Omega_{l|k]}\nn\\
&=&\nabla_i\Omega_{jk}+\frac13\big(2\nabla_i\Omega_{kj}+\nabla_{[j}\Omega_{k]i}\big)=\frac13\nabla_i\Omega_{jk}+\textrm{cyc}\nn\\
&=&\frac13(d\Omega)_{ijk}=0~.\nn\eea
Note that in the last step, the torsionless condition for $\Gamma$ is used to convert covariant derivatives to ordinary derivatives.

We shall drop the  $~\tilde{}~$  henceforth and perform the exponential map using the new connection\footnote{The use of such a connection
 for the RW model was previously suggested also in \cite{Rozansky:1996bq} and \cite{Thompson:1998vx}.}.
To obtain the Maurer-Cartan equation, we need to pull back $\bar\partial$.
The procedure is in \cite{Qiu:2009rf}, one tries to find a $\Theta$ such that
\bea \exp_{\xi}\big(\bar\partial+\{\Theta,\cdot\}\big)\exp^{-1}_{\xi}=\bar\partial~.\nn\eea
The left hand side can be expanded by using the formula
\bea
&&\exp{(-\xi\cdot\nabla)}{\cal O}\exp{\xi\cdot\nabla}={\cal O}
-\big[\xi\cdot\nabla,{\cal O}\big]+
\frac12\big[\xi\cdot\nabla,\big[\xi\cdot\nabla,{\cal O}\big]\big]+\cdots~,\nn\\
&&\xi\cdot\nabla=\xi^i\partial_{\textsl{x}^i}-\xi^i\xi^j\Gamma^k_{ij}\partial_{\xi^k}~,\nn\eea
where $\partial_{\textsl{x}^i}$ denotes the derivative with respect to  the base point of the expansion.

Suppose that the curvature tensor of $\Gamma$ is (1,1), then this infinite sum can be worked out explicitly, and one obtains (\ref{def_Theta}). That $\Theta$ satisfies the MC equation follows from the construction. In the general case, we cannot argue that the curvature of $\Gamma$ is of type (1,1) (this would have been true had we used the unique Hermitian connection associated with a Hermitian metric). As a result, starting from order $\xi^4$, there will be a new term $\xi^{l_1}\xi^{l_2}\xi^{l_3}\xi^{l_4}\big(R_{\bar il_1~l_2}^{\,~~k}R_{kl_3~l_4}^{~~~i}\big)\nabla_i$ in the expansion of $\exp{(-\xi\cdot\nabla)}\bar\partial\exp{\xi\cdot\nabla}$.
 Due to this complication, we cannot work out the exponential map to all orders as in (\ref{def_Theta}). However, as far as the perturbation theory is concerned, we only require the knowledge of $\Theta_{\bar i}$ up to order $\xi^3$, and the 3-valent vertex remains the same
\bea V_{ijk}=v^{\bar i}R_{\bar ii~k}^{~~l}\Omega_{lj}~,\label{RW_sym_hol}\eea
 which is automatically symmetric in $i, j, k$ if $\nabla$ preserves $\Omega$.
The MC equation up to this order is satisfied trivially $\bar\partial V=0$; the rest of the story of the perturbation theory follows through, and in conclusion, we are able to construct RW classes for a holomorphic symplectic manifold. The choice of the connection, since such connection is not unique, will not affect the RW class using a path integral
argument. In fact, this follows rather trivially from the fact that the vertex function is $\bar\partial$-closed: $\bar\partial_{[\bar i}R_{\bar j]j~l}^{\,~~k}=\nabla_{[\bar i}R_{\bar j]j~l}^{\,~~k}=\partial_jR_{\bar i\bar j~l}^{~~k}=0$, and that its variation is $\bar\partial$-exact: $\delta R_{\bar ij~l}^{~~k}=\bar\partial_{\bar i}\delta\Gamma_{jl}^k$. Therefore we see that under a change of connection the RW \emph{class} does not change.

Similar logic can be applied to the BF-RW, though in this case things work out quite nontrivially. As said above, the exponential map cannot be worked out to all orders. To order $\xi^2$,
\bea &&\exp^*\frac12\Omega_{ij}\delta X^i\delta X^j=\frac12\Omega_{ij}\delta\xi^i\wedge\delta\xi^j-\frac12\delta X^{\bar i}\wedge\delta\xi^j(R_{\bar im~n}^{~~i}\xi^m\xi^n)\nn\\
&&\hspace{6.2cm}
-\frac16\delta\xi^i\wedge\delta\xi^j(R_{im~n}^{~~q}\Omega_{qj}\xi^m\xi^n)+{\cal O}(\xi^3)~.\nn\eea
We emphasize that the $\Omega$ appearing here is a constant $\Omega(\textsl{x}_0)$.
We have to bring the symplectic form to canonical form, namely $\omega=\delta p_{\bar i}\delta X^{\bar i}+1/2\Omega_{ij}\delta\xi^i\wedge\delta\xi^j+\delta q_{\bar i}\delta v^{\bar i}$.
We already know how to reshuffle the second term: up to order $\xi^2$, it can be absorbed by a shift
\bea \tilde p_{\bar i}=p_{\bar i}+\frac16R_{\bar ii~k}^{~~l}\Omega_{lj}\xi^i\xi^j\xi^k+{\cal O}(\xi^3)~,\nn\eea
this is the same shift we have seen in the RW model. Now for the third term, it can be written as
\bea
-\frac16\delta\xi^i\wedge\delta\xi^j(R_{im~n}^{~~\,q}\Omega_{qj}\xi^m\xi^n)=
\frac{1}{24}\delta(R_{ij~l}^{~\,n}\Omega_{nk}\xi^j\xi^k\xi^l)\wedge\delta\xi^i~.\nn\eea
We therefore make a shift of $\xi$
\bea \tilde\xi^i=\xi^i-\frac{1}{24}\Omega^{ip}R_{pj~l}^{~~n}\Omega_{nk}\xi^j\xi^k\xi^l+{\cal O}(\xi^4)~.\nn\eea
Let us investigate the consequence of the second shift. After the exponential map and the first shift, the interaction term becomes
\bea &&-p_{\bar i}v^{\bar i}+A^A\mu_A+\frac12f_{AB}^{~~C}A^AA^BB_C=-\tilde p_{\bar i}v^{\bar i}+v^{\bar i}\frac16R_{\bar ii~k}^{~~l}\Omega_{lj}\xi^i\xi^j\xi^k+\frac12f_{AB}^{~~C}A^AA^BB_C\nn\\
&&\hspace{2cm}+A^A\Big(\mu_A(\textsl{x}_0)+(\partial_i\mu_A)\xi^i+\frac12(\nabla_i\partial_j\mu_A)\xi^i\xi^j
+\frac16(\nabla_i\nabla_j\partial_k\mu_A)\xi^i\xi^j\xi^k\Big)+{\cal O}(\xi^4)~.\nn\eea
We drop the tilde on $p_{\bar i}$ as usual. The second shift only affects the second term in the round brace, since the shift is of order $\xi^3$ and we are working up to order $\xi^3$.
So we get a new term 
\bea(\partial_i\mu_A)\xi^i\to (\partial_i\mu_A)\tilde\xi^i+\frac{1}{24}k_A^iR_{ij~l}^{~~n}\Omega_{nk}\tilde\xi^j\tilde\xi^k\tilde\xi^l\label{extra_term}\eea
for our 3-valent vertex. This unexpected term is in fact crucial for the connection independence of the theory.
To see this, suppose we choose a different connection (still preserving $J,\Omega$) $\Gamma\to\Gamma+\delta\Gamma$, then the variation of the interaction term of the action with respect to  the connection is\footnote{Since $\Psi$ is at least cubic, this is why the 3-valent vertex for the pure RW model (\ref{RW_sym_hol}) is not affected.}
\bea &&\delta_{\Gamma}S_{int}=\{\Psi,S_{int}\}~,\label{canonical_trans}\\
&&\Psi=-\frac16(\delta\Gamma^i_{mn}\Omega_{ip})\tilde\xi^m \tilde\xi^n \tilde\xi^p-\frac{1}{24}(\nabla_m\delta\Gamma^i_{nq})\Omega_{ip}
\tilde\xi^m \tilde\xi^n \tilde\xi^p\tilde\xi^q~.\nn\eea
The fact that $\delta_{\Gamma}S_{int}$ can be written as a canonical transformation should not be surprising. Because using either $\Gamma$ or $\Gamma+\delta\Gamma$ for the expansion, the original symplectic form $\delta X^i\Omega_{ij}\delta X^j$ is pulled back to the same $\delta\tilde\xi^i\Omega_{ij}(\textsl{x}_0)\delta\tilde\xi^j$, implying the existence of a canonical transformation. Note that (\ref{canonical_trans})
  fails without the extra term in (\ref{extra_term}).

Let us proceed to the analysis of the MC equation as we did in (\ref{Maure_Cartan}). If (\ref{Maure_Cartan}) stood, then our partition function would be $Q$-closed, and (\ref{canonical_trans}) would guarantee the connection independence. However, in this case (\ref{globalization}) fails due to the fact $R_{ij~l}^{~\;k}\neq0$. But having (\ref{globalization}) at hand is very desirable because it allows us to interpret our partition function as $Q$-cohomology classes. Here we present a simple solution to this problem. We see that the failure of (\ref{globalization}) is a mismatch between $Q\Theta^3$ and $\{\Theta_1,\Theta^4\}$ (using the notation of that section), we can fix this mismatch by hand through modifying $\Theta$ order by order. It turns out that in our case the required modification at order $\tilde\xi^3$ is an extra $1/8A^Ak_A^iR_{ij~l}^{~~n}\Omega_{nk}\tilde\xi^j\tilde\xi^k\tilde\xi^l$.

To conclude, for the BF-RW model, the modified 3-valent vertex is
\bea V=v^{\bar i}\Theta_{\bar i3}+A^A\mathsf{M}_{A3}+A^AU_A~;~~~U_A=\frac16k_A^iR_{ij~l}^{~\,n}\Omega_{nk}\tilde\xi^j\tilde\xi^k\tilde\xi^l~,\nn\eea
where $\Theta_{\bar i3},\SF{M}_{A3}$ are defined in (\ref{def_Theta}) and (\ref{def_M_A}), but now with $\tilde{\xi}$ in place of $\xi$. We find that, after quite a nasty bit of calculation, up to order $\tilde\xi^3$, the vertex $V$ does satisfy the MC equation,
\bea QV+\{A^A\mathsf{M}_{A2},V\}=0,\nn\eea
where $Q$ is as in (\ref{Q_BF}) and the bracket $\{~,~\}$ is in $\tilde\xi$-direction with respect to constant $\Omega(\textsl{x}_0)$.

To verify the connection independence, we vary $V$ with respect to the connection $\Gamma$,
\bea \delta_{\Gamma}V=\Big(\bar\partial+\{A^A\mathsf{M}_{A2},\cdot\}+A^Ak^l_A\nabla_l\Big)W=QW+\{A^A\mathsf{M}_{A2},W\}~,
~~~W=\frac16\delta\Gamma^n_{ik}\Omega_{nj}\xi^i\xi^j\xi^k~.\nn\eea
The variation of the BF-RW class is
\bea \delta_{\Gamma}c^q(\underbrace{V,\cdots V}_{q+1})=(q+1)c^q(QW+\{A^A\mathsf{M}_{A2},W\},\underbrace{V,\cdots V}_q)~,\nn\eea
where $c^q$ is the cocyle given by the path integral.
\bea
\delta_{\Gamma}c^q(V,\cdots V)&\sim&-q(q+1)c^q(W,QV,V,\cdots V)
+(q+1)c^q(\{A^A\mathsf{M}_{A2},W\},V,\cdots V)\nn\\
&=&q(q+1)c^q(W,\{A^A\mathsf{M}_{A2},V\},V,\cdots V)
+(q+1)c^q(\{A^A\mathsf{M}_{A2},W\},V,\cdots V)~,\nn\eea
where $\sim$ means the dropping of a $Q$-exact term. In the end, we see that the remaining terms combine to become an overall $sp(2n)$ rotation, and hence vanish.

In summary, we have given the prescription of constructing the equivariant RW-class associated with a holomorphic symplectic manifold, and demonstrated its invariance under the choice of connection. We believe that the way we fixed the problem above should originate from some deeper physical (mathematical) principles, which are probably related to the globalization issues discussed in \cite{Baulieu:2001fi} and the application of the Fedosov connection for handling perturbation theory on curved manifolds \cite{2001PThPS.144...38C}. With all these said, our result as it stands is nonetheless rigorous and impregnable.

\section{Summary and Outlook}
\label{end}

We have in this paper provided a systematic construction of the RW,  BF-RW and CS-RW models. We have shown that by applying the exponential map we can use a simple superfield computation to get exactly the same results as in the component approach. As an application of these theories, we showed that the evaluation of the partition function at each order of $\hbar$ is none other than applying the Chern-Weil homomorphism. This way, we constructed certain characteristic classes as a generalization of the original Rozansky-Witten classes.

There are a few interesting points in the final form of the characteristic classes. Firstly, looking at (\ref{ch_class_BFRW}), we observe the moment map does not appear un-differentiated. Thus these classes can be expressed using the vector field $k^i_A$ instead.  Thus the vertex
 looks as follows
 \beq
  V_{jkl}=dz^{\bi}R_{\bi j~l}^{~~n}\Omega_{nk}+A^{A} (\nabla_{j}\nabla_{k} k^n_{A}) \Omega_{nl} +  A^Ak_A^i R_{ij~l}^{~\,n} \Omega_{nk}~,
 \eeq{dkdke3303}
 where symmetrization in $j, k, l$ is assumed.
This suggests that our result is also valid for a group action that preserves $\Omega$, not necessarily Hamiltonian. Thus our result is applicable to much wider situations. Secondly, as we mentioned in the text, when we use $S^3$ as our source manifold, there exists the brute force gauge fixing that reduces all Feynman diagrams to only 3-valent. The cochains arising from such Feynman diagrams are automatically closed. On the other hand we know that there exist CE cocycles that are not represented by 3-valent graphs, so using AKSZ TFT on $S^3$, we will not reach any of the higher (and perhaps more interesting) classes. Yet, in our construction of these characteristic classes,
the 3-manifold plays an entirely passive role. In general one is allowed to replace the de Rham complex on the 3-manifold with any differential graded Frobenius algebra (see \cite{Cattaneo:2008ph, Bonechi:2009kx} for the construction)
  and the BV argument for the cocycle condition goes through just the same. It is therefore interesting to investigate this possibility and produce some higher classes, whose closedness is more genuinely dependent upon the path integral construction.

Last but not least, we provided an explicit formulae for the equivariant BF-RW classes on a holomorphic symplectic manifold. A physics based argument leads us to a vertex with a somewhat unexpected extra term, the existence of which is crucial for the invariance of the characteristic classes.
\bigskip\bigskip

\noindent{\bf\Large Acknowledgement}:
\bigskip

\noindent It is our pleasure to thank  Francesco Bonechi, Alberto Cattaneo, Andrew Dancer, Andrei Losev and
  Pavel Mn\"ev
  for discussions on this and related subjects.
  J.K. and M.Z. thank KITP, Santa Barbara where part of this work
was carried out.   We thank
the program "Higher Structures in Mathematics and Physics"
at the Erwin Schr\"odinger International Institute for Mathematical Physics where part of
this work was carried out.
The research of M.Z. was supported by VR-grant 621-2008-4273.
The research of M.Z. was supported in part by DARPA under Grant No. HR0011-09-1-0015 and
 by the National Science Foundation under Grant No. PHY05-51164.

\bigskip\bigskip

\bibliographystyle{utphys}

\bibliography{AKSZ}

\end{document}